\documentclass[lettersize,journal]{IEEEtran}
\IEEEoverridecommandlockouts

\usepackage{cite}
\usepackage{amsmath,amssymb,amsfonts}
\usepackage{mathtools}
\usepackage{amsthm}
\usepackage{algorithm}
\usepackage{algpseudocode}
\usepackage{etoolbox}
\usepackage{graphicx,color}
\usepackage{textcomp}
\usepackage[dvipsnames,rgb]{xcolor}
\usepackage{makecell}
\usepackage{colortbl}
\usepackage{float}
\usepackage{tikz}
\usetikzlibrary{arrows.meta, positioning, shapes.geometric,decorations.pathreplacing}
\usetikzlibrary{patterns}
\usepackage{siunitx}
\usepackage{tabularx}
\usepackage{multirow}
\usepackage{array}
\usepackage{acronym}
\usepackage{subcaption}
\usepackage{booktabs}
\usepackage{hyperref} 
\usepackage[capitalize,noabbrev]{cleveref}
\usepackage[textsize=tiny]{todonotes}
\usepackage{latexsym}
\usepackage{gensymb}
\usepackage{enumitem}
\usepackage{pdfpages}
\usepackage{setspace}
\usepackage{esvect}
\usepackage{comment}
\usepackage[normalem]{ulem}
\usepackage{soul}
\usepackage[most]{tcolorbox}
\usepackage[protrusion=true,expansion=true]{microtype}

\input{defs/macros}
\theoremstyle{plain}
\newtheorem{theorem}{Theorem}
\newtheorem{proposition}[theorem]{Proposition}
\newtheorem{lemma}[theorem]{Lemma}
\newtheorem{corollary}[theorem]{Corollary}
\theoremstyle{definition}

\newtheorem{assumption}[theorem]{Assumption}
\theoremstyle{remark}
\newtheorem{remark}[theorem]{Remark}

\definecolor{nyuviolet}{RGB}{87, 6, 140}
\long\def\revision#1{\begingroup\color{black}#1\endgroup}

\def\BibTeX{{\rm B\kern-.05em{\sc i\kern-.025em b}\kern-.08em
    T\kern-.1667em\lower.7ex\hbox{E}\kern-.125emX}}

\definecolor{revhdr}{RGB}{33, 85, 150}      
\definecolor{revbg}{RGB}{235, 243, 252}     
\definecolor{revtext}{RGB}{0, 0, 80}        
\definecolor{resphdr}{RGB}{0, 110, 85}      
\definecolor{respbg}{RGB}{235, 248, 243}    
\definecolor{resptext}{RGB}{0, 60, 40}      

\newtcolorbox{commentbox}[2][]{enhanced,
                            breakable,
                            colback=revbg,
                            colframe=revhdr,
                            coltitle=white,
                            coltext=revtext,
                            title={#2},
                            fonttitle=\bfseries,
                            boxrule=0.8pt,
                            arc=2pt,
                            left=4pt, right=4pt, top=2pt, bottom=2pt,
                            #1}
\newtcolorbox{responsebox}{enhanced,
                          breakable,
                          colback=respbg,
                          colframe=resphdr,
                          coltitle=white,
                          coltext=resptext,
                          title=Author Response,
                          fonttitle=\bfseries,
                          boxrule=0.8pt,
                          arc=2pt,
                          left=4pt, right=4pt, top=2pt, bottom=2pt,
                          }
\newtcolorbox{generalbox}[2][]{enhanced,
                            breakable,
                            colback=respbg,
                            colframe=resphdr,
                            coltitle=white,
                            coltext=resptext,
                            title={#2},
                            fonttitle=\bfseries,
                            boxrule=0.8pt,
                            arc=2pt,
                            left=4pt, right=4pt, top=2pt, bottom=2pt,
                            #1}

\definecolor{revisedhdr}{RGB}{150, 90, 0}       
\definecolor{revisedbg}{RGB}{253, 246, 233}     
\definecolor{revisedtext}{RGB}{60, 35, 0}       

\newtcolorbox{revisedbox}[1][]{enhanced,
                            breakable,
                            colback=revisedbg,
                            colframe=revisedhdr,
                            coltitle=white,
                            coltext=revisedtext,
                            title={Revised manuscript text\ifstrempty{#1}{}{ --- #1}},
                            fonttitle=\bfseries\small,
                            fontupper=\small,
                            boxrule=0.8pt,
                            arc=2pt,
                            left=4pt, right=4pt, top=2pt, bottom=2pt,
                            }

\AtBeginDocument{%
    \setlength\abovedisplayskip{2pt plus 1pt minus 1pt}%
    \setlength\belowdisplayskip{2pt plus 1pt minus 1pt}%
    \setlength\abovedisplayshortskip{1pt plus 1pt}%
    \setlength\belowdisplayshortskip{2pt plus 1pt minus 1pt}%
}

\AtBeginDocument{%
    \setlength\floatsep{4pt plus 2pt minus 2pt}%
    \setlength\textfloatsep{7pt plus 2pt minus 2pt}%
    \setlength\intextsep{4pt plus 2pt minus 2pt}%
    \setlength\dblfloatsep{4pt plus 2pt minus 2pt}%
    \setlength\dbltextfloatsep{7pt plus 2pt minus 2pt}%
    \setlength\abovecaptionskip{2pt}%
    \setlength\belowcaptionskip{0pt}%
    \renewcommand{\arraystretch}{0.95}%
}
\setlist{topsep=1pt,itemsep=0pt,parsep=0pt,partopsep=0pt}
\newacro{ACDD}{Alamouti with cyclic delay diversity}
\newacro{URLLC}{ultra-reliable low-latency communications}
\newacro{3GPP}{third generation partnership project}
\newacro{PHY}{physical layer}
\newacro{MIMO}{multiple-input multiple-output}
\newacro{MU-MIMO}{multi-user multiple-input multiple-output}
\newacro{SIMO}{single-input multiple-output}
\newacro{MISO}{multiple-input single-output}
\newacro{SISO}{single-input single-output}
\newacro{MRC}{maximum-ratio combining}
\newacro{SNR}{signal-to-noise ratio}
\newacro{CP}{cyclic prefix}
\newacro{CDD}{cyclic delay diversity}
\newacro{FSC}{frequency-selective channel}
\newacro{STC}{space-time coding}
\newacro{FFT}{fast Fourier transform}
\newacro{LMMSE}{linear minimum mean-square error}
\newacro{CFAC}{cross frequency AoA consistency}
\newacro{FER}{frame error rate}
\newacro{OFDM}{orthogonal frequency division multiplexing}
\newacro{OCDM}{orthogonal chirp division multiplexing}
\newacro{RMS}{root mean square}
\newacro{DS}{delay spread}
\newacro{FSC}{frequency-selective channel}
\newacro{CSI}{channel state information}
\newacro{LMMSE-PIC}{linear minimum mean squared error with parallel interference cancellation}
\newacro{PFE}{perfect-feedback equalizer}
\newacro{FD}{full-duplex}
\newacro{PDP}{power delay profile}
\newacro{PDF}{probability density function}
\newacro{DFT}{discrete Fourier transform}
\newacro{SDFT}{sparse DFT}
\newacro{ICI}{inter-carrier interference}
\newacro{OTFS}{orthogonal time frequency space}
\newacro{AWGN}{additive white Gaussian noise}
\newacro{SWH}{sparse Walsh-Hadamard}
\newacro{LLR}{log-likelihood ratio}
\newacro{PMF}{probability mass function}
\newacro{CRC}{cyclic redundancy check}
\newacro{PAM}{pulse amplitude modulation}
\newacro{QAM}{quadrature amplitude modulation}
\newacro{FWHT}{fast Walsh-Hadamard transform}
\newacro{MAP}{maximum a-posteriori}
\newacro{SC}{specular component}
\newacro{CFO}{carrier frequency offset}
\newacro{ISI}{inter-symbol interference}
\newacro{ZP}{zero-padding}
\newacro{EVD}{eigenvalue decomposition}
\newacro{BCJR}{Bahl, Cocke, Jelinek, and Raviv}
\newacro{WHT}{Walsh-Hadamard transform}
\newacro{APP}{a-posteriori probability}
\newacro{SILE-EPIC}{self-iterated linear equalizer with expectation propagation}
\newacro{EP}{expectation propagation}
\newacro{i.i.d.}{independent and identically distributed}
\newacro{CWCU}{component wise conditionally unbiased}
\newacro{MSE}{mean-square error}
\newacro{EXIT}{extrinsic information transfer}
\newacro{MI}{mutual information}
\newacro{PAPR}{peak-to-average power ratio}
\newacro{DFT-s}{discrete Fourier transform-spread}
\newacro{AMP}{approximate message passing}
\newacro{GAMP}{generalized \ac{AMP}}
\newacro{VAMP}{vector \ac{AMP}}
\newacro{RSC}{recursive systematic convolutional}
\newacro{QPSK}{quadrature phase-shift keying}
\newacro{CFAR}{constant false alarm rate}
\newacro{PD}{probability of detection}
\newacro{PFA}{probability of false alarm}
\newacro{RV}{random variable}
\newacro{CDF}{cumulative distribution function}
\newacro{HD-ZP}{half-duplex ZP}
\newacro{FD-CP}{full-duplex ZP}
\newacro{DFRC}{dual-function radar communication}
\newacro{SINR}{signal-to-interference-plus-noise ratio}
\newacro{ISAC}{integrated sensing and communication}
\newacro{SI}{self-interference}
\newacro{RSI}{residual self-interference}
\newacro{ADC}{analog-to-digital converter}
\newacro{DAC}{digital-to-analog converter}
\newacro{ED}{energy-detection}
\newacro{IDFT}{inverse discrete Fourier Transform}
\newacro{SFFT}{symplectic finite Fourier transform }
\newacro{CRB}{Cram{\'{e}}r-Rao bound}
\newacro{ZC}{Zadoff-Chu}
\newacro{RMSE}{root mean square error}
\newacro{MMSE}{minimum mean-square error}
\newacro{UW}{unique word}
\newacro{GFDM}{generalized frequency division multiplexing}
\newacro{RRC}{root-raised cosine}
\newacro{UB}{upper bound}
\newacro{CEF}{channel estimation field}
\newacro{TRX}{transceiver}
\newacro{IF}{intermediate frequency}
\newacro{RF}{radio-frequency}
\newacro{FPGA}{field programmable gate arrays}
\newacro{SDR}{software-defined radio}
\newacro{UWB}{ultra wideband}
\newacro{FR1}{frequency range 1}
\newacro{FR2}{frequency range 2}
\newacro{FR3}{frequency range 3}
\newacro{PCB}{printed circuit board}
\newacro{SMA}{SubMiniature version A}
\newacro{MUSIC}{multiple signal classification}
\newacro{CIR}{channel impulse response}
\newacro{FR}{Frequency Range}
\newacro{mmWave}{millimeter-wave}
\newacro{LoS}{line-of-sight}
\newacro{NLoS}{non-line-of-sight}
\newacro{AoD}{angle-of-departure}
\newacro{ESNR}{estimation SNR}
\newacro{AoA}{angle-of-arrival}
\newacro{SDNR}{signal-to-DMC-noise ratio}
\newacro{ULA}{uniform linear array}
\newacro{DMC}{dense multipath component}
\newacro{ML}{maximum-likelihood}
\newacro{IFFT}{inverse fast Fourier transform}
\newacro{LM}{Levenberg-Marquardt}
\newacro{ACF}{autocorrelation function}
\newacro{UWB}{ultra-wideband}
\newacro{SLAM}{simultaneous localization and mapping}
\newacro{STO}{sampling time offset}
\newacro{GLRT}{generalized likelihood ratio test}
\newacro{FD-ED}{frequency-domain energy detectors}
\newacro{IOU}{intersection over union}
\newacro{FLOP}{floating point operation}
\newacro{FLOPS}{floating point operations per second}
\newacro{LTBF}{long-term beamforming}
\newacro{UE}{user equipment}
\newacro{SRS}{sounding reference signal}
\newacro{BW}{bandwidth}
\newacro{RB}{resource block}
\newacro{RE}{resource element}
\newacro{BS}{base station}
\newacro{CG}{conjugate gradient}
\newacro{SRAM}{static random access memory}
\newacro{LVT}{low threshold voltage}
\newacro{HLS}{high-level synthesis}
\newacro{BLAS}{basic linear algebra subprograms}
\newacro{DSP}{digital signal processor}
\newacro{VLSI}{very large-scale integration}
\newacro{NR}{New Radio}
\newacro{ASIC}{application-specific integrated circuit}
\newacro{gNB}{next-generation NodeB}
\newacro{SCS}{subcarrier spacing}
\newacro{WMMSE}{weighted minimum mean-square error}
\newacro{WSR}{weighted sum-rate}
\newacro{ZF}{zero-forcing}
\newacro{UMB}{upper mid-band}
\newacro{DM-RS}{demodulation reference signal}
\newacro{XL-MIMO}{extremely large-scale \ac{MIMO}}
\newacro{LIS}{large intelligent surface}
\newacro{UM-MIMO}{ultra-massive \ac{MIMO}}
\newacro{THz}{terahertz}
\newacro{JSDM}{joint spatial division and multiplexing}
\newacro{CDM}{code-division multiplexing}
\newacro{PUSCH}{physical uplink shared channel}
\newacro{SVD}{singular value decomposition}
\newacro{CMAC}{complex multiply-accumulate operation}

\begin{document}

\title{Scalable Long-Term Beamforming \\
for Massive Multi-User MIMO
}

\author{
Ali Rasteh$^*$,
Amirreza Kiani$^{\diamond}$,
Marco Mezzavilla$^{\diamond}$,
and Sundeep Rangan$^*$ \\
	$^*$NYU WIRELESS, NYU Tandon School of Engineering, New York, USA \\
    $^\diamond$Dipartimento di Elettronica, Informazione e Bioingegneria (DEIB), Politecnico di Milano, Milan, Italy\\
	Email: ar7655@nyu.edu,
    amirreza.kiani@mail.polimi.it,
    marco.mezzavilla@polimi.it, 
    srangan@nyu.edu
}

\maketitle

\thispagestyle{empty}
\pagestyle{empty}

\acresetall
\begin{abstract}
Fully digital massive \revision{\ac{MIMO}} systems with large numbers (1000+) of antennas offer
capacity gains from spatial multiplexing and
beamforming, but receivers that scale to these array dimensions face
challenges in both channel estimation overhead and digital
computation.  Long-term beamforming addresses both by projecting the data onto
a low-dimensional subspace that can be tracked at a slow time scale from the
long-term channel parameters.  In this setting, we show how to
\revision{compute, in closed form, the projection matrix that maximizes a
capacity upper bound, using a matrix inverse square root; the same projection is
shown to maximize the mean post-projection \ac{SINR} exactly}.  Computationally efficient methods are then presented
for the matrix computation, realizable with matrix--matrix multiplies and hence
amenable to systolic array implementations in hardware.  \revision{Bounds}
on the \ac{SINR} degradation are derived, and ray tracing simulations in a realistic
rural uplink setting show \revision{a small loss relative to instantaneous \ac{MMSE} beamforming when the covariance is accurately estimated}.  \revision{The efficient Gram-domain form of the instantaneous \ac{MMSE}
receiver applies the maximum-ratio combining reduction before an inverse whose
dimension is the total number of streams.  Against this baseline, the method
estimates and refreshes beamforming coefficients three orders of magnitude less
often and decouples the real-time path across users.  With the conjugate-gradient
solve and a rank-one projection, its total arithmetic cost is $2\%$ higher at the
ten-user operating point and lower above a stream-dimension crossover
that we characterize.}
\end{abstract}


\acresetall

\section{Introduction}
\label{sec:intro}

\IEEEPARstart{M}{assive} \ac{MIMO} {\cite{marzetta2016fundamentals}},
where the base stations use a large number of antenna elements and streams,
\revision{was a key technology for increasing 5G capacity}.  There is now considerable interest in expanding the \ac{MIMO} antenna dimensions further.  For example, the simulation study \cite{nokia2025massiveMIMO}
shows that \ac{MIMO} systems with 1024
antenna elements (at least five times \revision{the antenna count of} current commercial base stations) can increase spectral efficiency by at least fourfold.  Such massive \ac{MIMO} systems, sometimes called \emph{extreme \ac{MIMO}}, are particularly valuable in the emerging upper mid-band 
\cite{kang2024cellular,nokia2025coverage7to15GHz}.  An end-to-end cellular link has been demonstrated in this band \cite{mezzavilla2025prototyping}.  \revision{High-dimensional arrays also improve}
interference cancellation \cite{jia2025joint}.

Implementing massive \ac{MIMO} systems at these scales \revision{raises two challenges} {\cite{dai2021scalable}}. 
\revision{The first is} the \textbf{channel estimation overhead}.  
\revision{From an information-theoretic perspective, at high \ac{SNR}}, where \ac{MIMO} has advantages, the channel estimation overhead typically scales linearly with the number of streams \cite{marzetta2002capacity}.  The pilot overhead for tracking small-scale fading across large numbers of streams, particularly in mobile environments, becomes overwhelming.
\revision{The second is} the \textbf{computational complexity}.
Theoretical \ac{MIMO} receivers based on the \ac{MSE} or \ac{ZF} criteria {\cite{lin2019hybrid}} require matrix inverses on the order of the number of streams and antennas. These matrices theoretically need to be re-computed in each coherence time-frequency block and each scheduling instance, and they can become computationally prohibitive with a large number of streams and antennas.

This paper addresses these challenges for a cellular uplink. Specifically, we consider a single base station with $N_{\subsf rx}$ receive antennas receiving data from $N_{\subsf UE}$ mobile \ac{UE} stations, each \ac{UE} with $N_s$ streams.
We consider a \ac{MU-MIMO} scenario where 
all the streams are to be received on the same time-frequency resources and, hence, must be spatially separated.

\subsection*{Contributions}
We present a low-overhead,
computationally efficient approach to large-scale uplink \ac{MU-MIMO}.
To overcome the cost of acquiring and refreshing full-dimensional instantaneous \ac{CSI}, we use the so-called \emph{long-term beamforming} \revision{(\acs{LTBF})}\acused{LTBF} of \cite{lozano2007long}.\revision{\ Section~\ref{sec:signalling} quantifies that cost in 5G \ac{NR} resources. Long-term beamforming reduces the rate at which beamforming coefficients must be produced and the accuracy demanded of each pilot; it does not reduce the number of pilot \acp{RE}: both receivers need the same \ac{DM-RS}, and long-term beamforming adds a small \ac{SRS} term.}  In long-term beamforming, we estimate a low-rank projection for each user.  This low-rank projection 
is stable over the \emph{large-scale} propagation parameters, such as angles of arrival and path gains, which vary slowly and can be estimated with minimal overhead. 
\revision{The contributions are as follows.}

\revision{First, we derive a closed-form low-rank projection that maximizes a
Jensen upper bound on the ergodic capacity (Lemma~\ref{lem:capupper}); we do
not claim that it maximizes the ergodic capacity itself.  The same projection
exactly maximizes the mean post-projection \ac{SINR} (Corollary~\ref{cor:sinr})
and any objective increasing in the projected signal-to-interference
eigenvalues, so the design does not depend on the choice of bound.}  Second, we give two efficient approximations of the matrix
inverse square root the projection requires, namely the \ac{CG} method and a
matrix polynomial expansion, built from matrix products rather than direct
inversion and hence far more parallelizable.  Third, we bound what these
approximations cost in performance. Specifically, 
\revision{Proposition~\ref{prop:subspace} bounds, without structural
assumptions, the subspace error of the computed projection by the solver
residual, and Proposition~\ref{prop:error} then gives} a lower bound on the
resulting \ac{SINR} degradation \revision{for a single-stream \ac{UE} with a
rank-one covariance}, showing that
higher-\ac{SINR} users need greater precision.  Fourth, we realize the scheme
with 5G-\ac{NR}-like signals (Sec.~\ref{sec:soln}), estimating the long-term
terms from periodic \ac{SRS} and tracking small-scale fading post-projection
with in-band \ac{DM-RS}.  Finally, we demonstrate it by ray tracing in a
realistic rural uplink, \revision{showing a small degradation when the
covariance is accurately estimated while beamforming coefficients are refreshed
three orders of magnitude less often than with instantaneous beamforming}.

\subsection*{Related Work}

The use of long-term beamforming to reduce channel estimation overhead is well known and dates back at least to \cite{lozano2007long}.
Several works have examined multi-user and hybrid analog–digital \ac{MIMO} systems, where only partial or low-dimensional \ac{CSI} is available 
\cite{visotsky2002space,jafar2004transmitter}.
\revision{Most of this work has} focused on the downlink
\cite{lozano2007long}. Comparatively fewer works have addressed scalable uplink multi-user processing using long-term statistics, especially in the context of extreme \ac{MIMO} arrays with hundreds or thousands of antennas. 
\revision{Closest in spirit is \ac{JSDM} \cite{adhikary2013joint,nam2014joint},
which partitions the \acp{UE} into groups with approximately common covariance
eigenspaces and block-diagonalizes across groups, so that instantaneous \ac{CSI}
is needed only within a group.  \ac{JSDM} is therefore interference-aware
between groups and, like the schemes above, energy-preserving within one.
It also requires a grouping step, which the projection of this paper does not.}

\revision{%
These families differ in two main respects: what \ac{CSI} the dimension-reducing
projection is built from, and how multi-user interference enters its design.
Instantaneous \ac{MMSE}/\ac{ZF} reception
\cite{heath2018foundations,jeon2019decentralized} builds no projection at all
and handles interference exactly and jointly, at the price of a beamformer
solve on the coherence time scale---$\mc{O}(N_{\subsf rx}^3)$
full-dimensionally, and $\mc{O}(N_{\subsf rx}(N_{\subsf UE}N_s)^2)$ in the
efficient Gram-domain realization of \cite{jeon2019decentralized} that
Sec.~\ref{sec:baseline} adopts as the baseline throughout.  Classical
statistical beamforming \cite{lozano2007long,visotsky2002space,jafar2004transmitter}
truncates a \ac{UE}'s own covariance $\bs{Q}_i$, which preserves that
\ac{UE}'s energy but ignores the other \acp{UE}' covariances, whereas
\ac{JSDM} \cite{adhikary2013joint,nam2014joint} adds interference awareness
only between groups. Hybrid analog--digital designs \cite{lin2019hybrid}
constrain the projection to the available \ac{RF} chains; and beamspace processing
\cite{sayeed2013beamspace,mirfarshbafan2020beamspace,dai2021scalable} selects
columns of a \ac{DFT} basis and relies on beam separation.  Statistical
beamforming and \ac{JSDM} reduce dimension at the same
$\mc{O}(rN_{\subsf rx})$ cost per resource element as the projection here, so
they differ only in the information from which the projection is built.
The step taken here is that \eqref{eq:Giopt} whitens by the aggregate
covariance $\bs{Q}$ before truncating, so the retained subspace is the one with
the largest signal-to-interference ratio rather than signal power, and
interference enters every \ac{UE}'s projection at the long-term time scale, not
only between groups.  Section~\ref{sec:ops} compares this step with
eigen-beamforming (the same projection without the whitener) and with \ac{DFT}
beamspace selection on the same covariance estimate, and identifies where the
whitener helps.
}

There is also a large body of work on efficient \ac{VLSI} implementations of \ac{MIMO}, most of it aimed at the instantaneous \ac{MMSE} detector itself through hardware-friendly approximate matrix inversion---Neumann series, conjugate gradient and Gauss--Seidel \cite{yin2015vlsi,wu2013approximate}---or through polynomial and Chebyshev matrix-function approximations \cite{kammoun2014linear}, all of which still process at the coherence time scale.  Our approach instead applies a polynomial approximation to the long-term spatial covariance, shifting the computational burden to a much slower time scale, and the low-rank projection of Lemma~\ref{lem:capupper} is novel independent of how it is implemented.

Some literature in the field relies on sparsity and beamspace processing
\cite{dai2021scalable,mirfarshbafan2020beamspace,sayeed2013beamspace}.  This work does not operate in beamspace\revision{.  A complementary beamspace change of basis that sparsifies the long-term system matrix and accelerates its inversion is developed separately in \cite{kiani2026lowrank}.}  One contribution of this work is to connect these methods to the requirements of \ac{MU-MIMO} equalization. In particular, we show that the distribution of eigenvalues is connected to both power control and the range of \acp{SINR}.

Another line of recent work uses model-driven deep learning (deep unfolding) to accelerate \ac{MU-MIMO} precoding and detection by unrolling iterative algorithms---a Riemannian \ac{CG} precoder for constant-envelope massive \ac{MU-MIMO} and unfolded \ac{WMMSE} beamformers---into trainable networks \cite{he2020model,hu2020iterative}\revision{.  Our scheme is} training-free and fully deterministic, retaining an explicit linear solver whose preconditioning and step-size rules are specified analytically from long-term channel statistics.

\revision{%
The array scales considered here also connect this work to three adjacent lines
of research.  \emph{\ac{XL-MIMO}} and \emph{\acp{LIS}} study apertures large
enough that a \ac{UE} lies in the radiative near field and different parts of
the array see different propagation
\cite{hu2018beyond,bjornson2020power,carvalho2020nonstationarities,wang2024xlmimo,lu2024nearfield,amiri2018extremely,cui2022channel}.  A measurement system for near-field path parameters in the upper mid-band is described in \cite{rasteh2024nearfield}.
\emph{\ac{UM-MIMO}} extends massive \ac{MIMO} into the \ac{mmWave} and \ac{THz} bands
\cite{akyildiz2016realizing,faisal2020ultramassive,sarieddeen2021overview}.
\emph{Gigantic \ac{MIMO}} is the corresponding evolution in the \ac{UMB}, the
regime targeted here: hundreds to thousands of elements at \SIrange{7}{15}{GHz},
where the array rather than the bandwidth supplies the capacity
\cite{bjornson2025gigantic,kang2024cellular,nokia2025coverage7to15GHz,bjornson2019massive}.
These lines motivate this work, because the per-coherence-block beamformer
computation it removes is what makes such arrays impractical with an
instantaneous receiver; they also limit its scope, because near-field focusing
and spatial non-stationarity both raise the rank of $\bs{Q}_i$ on which the
projection depends.  Our \SI{3.5}{GHz} far-field, spatially stationary
evaluation tests neither mechanism; Sec.~\ref{sec:rank} instead varies the
propagation richness directly and reports the rank the method needs.}

\section{Multi-User Long-Term Beamforming}  \label{sec:mimo_proc}

\revision{\subsection{System Model}\label{sec:sysmodel}}
We consider a multi-user \ac{MIMO} uplink 
where $N_{\subsf UE}$ \acp{UE} are transmitting
in a common time-frequency resource.
\revision{The model of this subsection is used throughout the paper, and
Appendix~\ref{app:notation} summarizes the notation.  Throughout, $(\cdot)\herm$
is the conjugate transpose and $\bs{Z}^{1/2}$ denotes the \emph{unique Hermitian
positive semidefinite} square root of a Hermitian positive semidefinite
$\bs{Z}$, so that $\bs{Z}^{\subsf H/2} = \bs{Z}^{1/2}$, $\bs{Z}^{-\subsf H/2} =
\bs{Z}^{-1/2}$ for $\bs{Z}\succ0$, and the factorization
$\bs{Z} = \bs{Z}^{1/2}\bs{Z}^{1/2}$ is simultaneously the congruence
$\bs{Z} = \bs{Z}^{\subsf H/2}\bs{Z}^{1/2}$.}
Suppose each \ac{UE} transmits $N_s$ streams.
The received vector at the base station
\revision{on subcarrier $n$ of \ac{OFDM} symbol $k$}
can be described \cite{heath2018foundations}
by:
\begin{equation} \label{eq:mumimo}
    \bs{y}[n,k] = \sum_{i=1}^{N_{\subsf UE}}\bs{H}_i[n,k]\bs{x}_i[n,k] + \bs{w}[n,k],
\end{equation}
\revision{where $\bs{y}[n,k] \in \C^{N_{\subsf rx}}$ is the received vector
across the $N_{\subsf rx}$ base-station antennas,
$\bs{H}_i[n,k] \in \C^{N_{\subsf rx}\times N_s}$ is the channel matrix of
\ac{UE} $i$, $\bs{x}_i[n,k] \in \C^{N_s}$ is its transmitted symbol vector, and
$\bs{w}[n,k] \sim \mc{CN}(\bs{0}, N_0\bs{I})$ is the receiver noise, independent
across $(n,k)$ and independent of the $\bs{x}_i[n,k]$.}
Note that $\bs{H}_i[n,k]$ includes any \revision{precoding applied} at
\ac{UE} $i$.
We let $\mc{E}_{x_i}$ denote the energy per \ac{UE} per symbol and assume
$\Exp[\bs{x}_i[n,k]\bs{x}_i\herm[n,k]] = (\mc{E}_{x_i}/N_s)\bs{I}$.
In linear \ac{MIMO} processing, the base station
will compute an estimate of $\bs{x}_i[n,k]$
by:
\begin{equation} \label{eq:xFy}
    \wh{\bs{x}}_i[n,k] = \bs{F}_i[n,k]\bs{y}[n,k],
\end{equation}
where $\bs{F}_i[n,k]$ is the so-called spatial
equalization matrix. Each spatial equalization matrix $\bs{F}_i$ will attempt to align with the desired signal $\bs{x}_i$ while nulling the other signals $\bs{x}_j$ for $j\neq i$.

\subsection{Baseline: Instantaneous MMSE Beamforming}
\label{sec:baseline}
If the channels $\bs{H}_i[n,k]$
were known at the base station receiver, then
the \emph{instantaneous} \ac{MMSE} receiver is given by:
\begin{equation} \label{eq:Finstant}
    \bs{F}_i[n,k] = \alpha\bs{H}_i[n,k]\herm\left( \bs{I} + \sum_{\revision{j}=1}^{N_{\subsf UE}} \alpha_{\revision{j}} \bs{H}_{\revision{j}}[n,k]\bs{H}_{\revision{j}}[n,k]\herm  \right)^{-1},
\end{equation}
\revision{where $\alpha > 0$ is an arbitrary scaling, common to every stream of
\ac{UE} $i$ and therefore without effect on any \ac{SINR} reported in this
paper.  Moreover, the constant}
\begin{equation} \label{eq:alphai_def}
    \alpha_i = \frac{\mc{E}_{x_i}}{N_0N_s}
\end{equation}
is the transmit \ac{SNR} \revision{of \ac{UE} $i$ per stream}.

\revision{%
Equation \eqref{eq:Finstant} is written with an
$N_{\subsf rx}\times N_{\subsf rx}$ inverse, but that is not its cheapest
realization.  Stack the per-\ac{UE} channels as
$\bs{H}[n,k] = [\bs{H}_1[n,k],\ldots,\bs{H}_{N_{\subsf UE}}[n,k]] \in
\C^{N_{\subsf rx}\times N_{\subsf UE}N_s}$, let
$\bs{\Lambda}_\alpha = \diag(\alpha_1\bs{I}_{N_s},\ldots,\alpha_{N_{\subsf UE}}\bs{I}_{N_s})$,
and let $\bs{J}_i$ select the $N_s$ columns of \ac{UE} $i$.  The push-through
identity then turns \eqref{eq:Finstant} into the algebraically identical
\emph{Gram-domain} form
\begin{equation} \label{eq:Fgram}
    \bs{F}_i[n,k] = \alpha\,\bs{J}_i\herm
    \left( \bs{I} + \bs{H}\herm[n,k]\bs{H}[n,k]\bs{\Lambda}_\alpha \right)^{-1}
    \bs{H}\herm[n,k],
\end{equation}
in which the information-lossless \ac{MRC} reduction $\bs{H}\herm\bs{y}$ is
applied first and an $N_{\subsf UE}N_s \times N_{\subsf UE}N_s$ matrix is
inverted~\cite{jeon2019decentralized}.  The two forms are the same receiver and give identical \acp{SINR}, but the
order of computation is different.  With the Gram-domain receiver the
beamformer computation falls from $\mc{O}(\chi_{\subsf I}N_{\subsf rx}^3)$ to
$\mc{O}(N_{\subsf rx}(N_{\subsf UE}N_s)^2 + \chi_{\subsf I}(N_{\subsf UE}N_s)^3)$
per coherence block.  Therefore in the remainder of the paper, ``instantaneous
\ac{MMSE}'' refers to \eqref{eq:Fgram} unless the full-dimensional form is named
explicitly, and Table~\ref{tab:complexity} reports both.}


\subsection{Long-Term Beamforming}
\label{sec:ltbf}
The conventional instantaneous \ac{MMSE} receiver maximizes the \ac{SINR}, but
\revision{both realizations \eqref{eq:Finstant} and \eqref{eq:Fgram}} require the channel
matrices $\bs{H}_i[n,k]$ of all \acp{UE} at all time-frequency points $(n,k)$
\revision{to be estimated, stored and turned into fresh combining weights once
per coherence block}.  For large numbers of \acp{UE}---the target of this work---\revision{estimating these channels and refreshing the combining
weights at that rate} is too expensive.

We thus follow a \emph{long-term beamforming} strategy
\cite{lozano2007long}.  To decode the symbols $\bs{x}_i[n,k]$ of \ac{UE} $i$,
rewrite \eqref{eq:mumimo} as:
\begin{equation} \label{eq:mumimov}
\begin{aligned}
    \bs{y}[n,k] &= \bs{H}_i[n,k]\bs{x}_i[n,k] + \sum_{j\neq i}\bs{v}_j[n,k] + \bs{w}[n,k],
    \\
    \bs{v}_j[n,k] &= \bs{H}_j[n,k]\bs{x}_j[n,k],
\end{aligned}
\end{equation}
where $\bs{v}_j$ is the interference from \ac{UE} $j$.  The key idea is to
project $\bs{y}[n,k]$ into a low-dimensional subspace that approximately nulls
the $\bs{v}_j[n,k]$:
\begin{equation} \label{eq:zproj}
    \bs{z}_i[n,k] = \bs{G}_i \bs{y}[n,k],
\end{equation}
where $\bs{G}_i$ is $r \times N_{\subsf rx}$ for some $r < N_{\subsf rx}$.  It
is held constant over a long period and across all subcarriers $n$, since it
relies on long-term spatial statistics that are stable over the bandwidth, and
consequently does not compensate for the frequency-selective small-scale
fading.  That is addressed in a second stage: the symbols are recovered from
$\bs{z}_i[n,k]$ by a low-dimensional, per-subcarrier instantaneous \ac{MMSE}
equalizer $\bs{W}_i[n,k]$,
\begin{equation} \label{eq:xhat_stage2}
\wh{\bs{x}}_i[n,k] = \bs{W}_i[n,k] \bs{z}_i[n,k],
\end{equation}
computed from the effective low-rank channel $\tilde{\bs{H}}_i[n,k] = \bs{G}_i
\bs{H}_i[n,k]$.
\revision{%
Splitting the beamforming operation into the two phases
\eqref{eq:zproj}--\eqref{eq:xhat_stage2} does not reduce the
per-resource-element arithmetic: applying $\bs{G}_i$ costs
$rN_{\subsf rx}$ operations per \ac{UE} per resource element in
\eqref{eq:zproj}, whereas the \ac{MRC} reduction in \eqref{eq:Fgram} takes
$N_{\subsf rx}N_s$, so the two differ only by the ratio $r/N_s$.  Each scheme
also carries a second, reduced-dimension term per resource element---$rN_s$
in \eqref{eq:xhat_stage2} and $(N_{\subsf UE}N_s)^2$ for \eqref{eq:Fgram}---and both are negligible against $N_{\subsf rx}$.  The advantage of the proposed
method is that $\bs{W}_i[n,k]$ is computed from an $r$-dimensional
effective channel rather than an $N_{\subsf rx}$-dimensional one, and
decoupled across \acp{UE}: the receiver inverts one $r\times r$
matrix per \ac{UE}, with no joint $N_{\subsf UE}N_s$-dimensional inverse and no exchange of
instantaneous \ac{CSI}.  The consequence, quantified in
Sec.~\ref{sec:complexity}, is that the only quantity refreshed at the coherence
time scale is an $r\times N_s$ effective channel per \ac{UE}, instead of the
$N_{\subsf rx}\times N_{\subsf UE}N_s$ matrix that \eqref{eq:Fgram} needs.}

\subsection{Optimizing the Projection Matrix}
\label{sec:optproj}
Following \cite{yu2020long}, we treat the channel matrix $\bs{H}_j[n,k]$ of
each \ac{UE} $j$ as random with spatial covariance
\begin{equation}   \label{eq:Qjdef}
    \bs{Q}_j := \Exp\left[ 
        \bs{H}_j[n,k]\bs{H}_j[n,k]\herm \right],
\end{equation}
the expectation being over a period in which the large-scale parameters remain
constant while the small-scale parameters vary.  Rewrite \eqref{eq:mumimov} as
$\bs{y} = \bs{H}_i\bs{x}_i + \bs{d}_i$, dropping $n,k$, where
$\bs{d}_i = \sum_{j \neq i} \bs{v}_j + \bs{w}$ is the interference plus noise.
Its covariance normalized by $N_0$, and the aggregate covariance, are
\begin{equation} \label{eq:Qdef}
    \bs{R}_i := \frac{1}{N_0}\Exp\left[ \bs{d}_i\bs{d}_i\herm \right] = \bs{I}
    + \sum_{j \neq i} \alpha_j \bs{Q}_j,
    \quad
    \bs{Q} := \bs{I} + \sum_{j=1}^{N_{\subsf UE}} \alpha_j \bs{Q}_j,
\end{equation}
so that $\bs{R}_i = \bs{Q} - \alpha_i \bs{Q}_i$.  The projection output
\eqref{eq:zproj} is then
\begin{equation} \label{eq:zprojd}
    \bs{z}_i = \tilde{\bs{H}}_i\bs{x}_i + \tilde{\bs{d}}_i,
    \qquad
    \tilde{\bs{H}}_i = \bs{G}_i \bs{H}_i,
    \quad
    \tilde{\bs{d}}_i = \bs{G}_i \bs{d}_i,
\end{equation}
whose projected interference-plus-noise covariance is
$\tilde{\bs{R}}_i := N_0^{-1}\Exp[ \tilde{\bs{d}}_i\tilde{\bs{d}}_i\herm ]
= \bs{G}_i\bs{R}_i\bs{G}_i\herm$.
The ergodic capacity of \eqref{eq:zprojd}, the expectation being over the
small-scale variation in $\bs{H}_i$, is $C_i(\bs{G}_i) = \Exp[ \log_2 \det(\bs{I} + \revision{\alpha_i}
\bs{G}_i\bs{H}_i \bs{H}_i\herm \bs{G}_i\herm \tilde{\bs{R}}_i^{-1} )]$, and by
Jensen's inequality $C_i(\bs{G}_i) \leq \wbar{C}_i(\bs{G}_i)$, where
\begin{align}
    \wbar{C}_i(\bs{G}_i) &:=  \log_2 \mathrm{det}(\bs{I} + \revision{\alpha_i}
    \bs{G}_i \Exp[\bs{H}_i \bs{H}_i\herm ] \bs{G}_i\herm \tilde{\bs{R}}_i^{-1} )
    \nonumber \\
    &= \log_2 \mathrm{det}(\bs{I} + \revision{\alpha_i}
    \Lambda_i(\bs{G}_i) ),
    \nonumber \\
    \Lambda_i(\bs{G}_i) &:=
        \bs{G}_i \bs{Q}_i \bs{G}_i\herm\tilde{\bs{R}}_i^{-1}.
    \label{eq:capupper}
\end{align}
The problem is to find the rank-$r$ projection
$\bs{G}_i \in \C^{r \times N_{\subsf rx}}$ maximizing $\wbar{C}_i(\bs{G}_i)$,
and the following lemma solves it.

\begin{lemma} \label{lem:capupper}
For a given projection rank $r$,
one matrix $\bs{G}_i$ that maximizes
\revision{the capacity upper bound} $\wbar{C}_i(\bs{G}_i)$ \revision{of
\eqref{eq:capupper}} is
\begin{equation} \label{eq:Giopt}
    \bs{G}_i = [\bs{Q}_i^{1/2}\bs{Q}^{-1/2}]_r\bs{Q}^{-1/2}
\end{equation}
where $[\bs{Z}]_r$ is the matrix with the
$r$ rows of the right singular vectors of $\bs{Z}$ for the $r$ largest singular values.
\revision{The maximizer is determined only up to a change of basis of its row
space ($\wbar{C}_i(\bs{T}\bs{G}_i) = \wbar{C}_i(\bs{G}_i)$ for every invertible
$\bs{T}\in\C^{r\times r}$).}
\end{lemma}
\begin{proof}
    See Appendix~\ref{sec:proof_projection}.
\end{proof}

\revision{%
Optimizing the \ac{SINR}, the metric in which our numerical results are
reported, yields the same projection, as the following bound-free argument
shows.

Take $N_s = 1$ and write $\bs{H}_i = \bs{h}_i$.  After the projection, the
\ac{MMSE} receiver applied to $\bs{z}_i = \tilde{\bs{h}}_i x_i +
\tilde{\bs{d}}_i$ achieves the \ac{SINR}
\begin{equation} \label{eq:sinr_proj}
    \gamma_i(\bs{G}_i)
    \;=\; \alpha_i\, \tilde{\bs{h}}_i\herm \tilde{\bs{R}}_i^{-1}
    \tilde{\bs{h}}_i,
    \qquad
    \tilde{\bs{h}}_i = \bs{G}_i\bs{h}_i,
\end{equation}
with $\tilde{\bs{R}}_i = \bs{G}_i\bs{R}_i\bs{G}_i\herm$ as in \eqref{eq:zprojd}.
Since $\tilde{\bs{R}}_i$ does not depend on the small-scale fading, the
expectation of \eqref{eq:sinr_proj} over that fading can be evaluated exactly.

\begin{corollary} \label{cor:sinr}
Let $N_s = 1$.  Then the mean post-projection \ac{SINR} is
\begin{equation} \label{eq:Egamma}
    \Exp\!\left[\gamma_i(\bs{G}_i)\right]
    \;=\; \alpha_i \tr\!\big(\Lambda_i(\bs{G}_i)\big)
    \;=\; \alpha_i \sum_{k=1}^{r} \frac{\mu_k}{1 - \alpha_i\mu_k},
\end{equation}
where $\mu_1,\ldots,\mu_r$ are the Ritz values of
$\bs{Q}_i\wtn = \bs{Q}^{-\subsf H/2}\bs{Q}_i\bs{Q}^{-1/2}$ on the row space of
$\bs{G}_i$.  Consequently $\Exp[\gamma_i]$ is maximized by the same
$\bs{G}_i$ as in \eqref{eq:Giopt}.
\end{corollary}
\begin{proof}
See Appendix~\ref{sec:proof_projection}.
\end{proof}

The two routes coincide because both objectives are functions of the same $r$
Ritz values $\mu_k$ and both are strictly increasing in each of them.  The
optimal subspace is thus the one that makes the Ritz values of $\bs{Q}_i\wtn$ as
large as possible, and by the Ky Fan/Courant--Fischer principle that subspace is
spanned by the $r$ principal eigenvectors of $\bs{Q}_i\wtn$---whichever
increasing objective is chosen.  Equation~\eqref{eq:Giopt} is therefore optimal for the
Jensen bound, for the mean \ac{SINR} and for any measure increasing in each
$\mu_k$, but it does not in general maximize the
ergodic capacity $C_i$ itself: on a sum-of-subpaths ensemble the Jensen bound is
$13$--$61\%$ loose, yet numerically maximizing $C_i$ improves on
\eqref{eq:Giopt} by at most $0.23\%$ over $1440$ instances.}

The lemma provides, in principle, a simple recipe for long-term multi-user
beamforming: estimate the spatial covariance matrices $\bs{Q}_j$ and form
$\bs{Q}$ from \eqref{eq:Qdef}; compute the projection $\bs{G}_i$ from
\eqref{eq:Giopt}; and apply it with \eqref{eq:zproj},
then demodulate and decode as a single-user system, treating interference as
noise.

\label{sec:challenges}Three challenges stand in the way of implementing this
strategy: estimating $\bs{Q}_i$, i.e., forming $\wh{\bs{Q}}_i$; computing
$\bs{Q}^{-1/2}$ in \eqref{eq:Giopt}; and forming $\bs{G}_i$ from
\eqref{eq:Giopt}, \revision{whose $N_{\subsf rx}\times N_{\subsf rx}$ products
must be kept off the per-resource-element path}.

\section{Proposed Solution}
\label{sec:soln}

\revision{This section addresses the three challenges listed at the end of
Sec.~\ref{sec:optproj}.}

\subsection{Estimation of the Spatial Covariance Matrices}
\label{sec:spatcovest}

The key to estimating the spatial covariance matrix $\bs{Q}_j$ in
\eqref{eq:Qjdef} is that the matrix is generally low rank, since there are
typically a limited number of dominant paths, so a limited number of
measurements suffices.  For the 5G uplink these come from the \ac{SRS}: once
per period $T_{\subsf LT}$, which we will call the \emph{long-term estimation
period}, each \ac{UE} sounds the band (Sec.~\ref{sec:signalling}), and the base
station forms $N_{\subsf SRS}$ narrowband estimates at different frequencies of
that sounding, the $m$th by correlating the received signal $\bs{y}[m]$ with
the known pilot sequence $\bs{x}^{\subsf SRS}_i[m]$ from \ac{UE} $i$. \revision{A \ac{UE} with $N_s$ transmit antennas sounds all of them, so the $m$-th estimate is an $N_{\subsf rx}\times N_s$ block $\wh{\bs{H}}_j[m]$, an estimate of $\bs{H}_j$ at that frequency.  The base station concatenates the $N_{\subsf SRS}$ blocks horizontally into
\begin{equation} \label{eq:Hhatdef}
    \wh{\bs{H}}_j = \left[\wh{\bs{H}}_j[1],\ldots,\wh{\bs{H}}_j[N_{\subsf SRS}]\right]
    \;\in\; \C^{N_{\subsf rx}\times \wbar{N}},
    \qquad
    \wbar{N} := N_sN_{\subsf SRS},
\end{equation}
and estimates the spatial covariance \eqref{eq:Qjdef} by the sample average
over the $N_{\subsf SRS}$ \emph{snapshots},}
\begin{equation} \label{eq:QHprod}
    \wh{\bs{Q}}_j = \frac{1}{N_{\subsf SRS}} \wh{\bs{H}}_j\wh{\bs{H}}_j\herm
    \revision{{}= \frac{1}{N_{\subsf SRS}}\sum_{m=1}^{N_{\subsf SRS}}
    \wh{\bs{H}}_j[m]\wh{\bs{H}}_j[m]\herm},
\end{equation}
\revision{%
held across the band and refreshed once per $T_{\subsf LT}$, on the same
grounds as $\bs{G}_i$ in Sec.~\ref{sec:ltbf}.  Since \eqref{eq:Qjdef}
sums the $N_s$ per-stream outer products, the normalization is by the
$N_{\subsf SRS}$ snapshots, not the $\wbar{N}$ columns, which would estimate
$\bs{Q}_j/N_s$ and understate the interference by that factor.  So normalized,
\eqref{eq:QHprod} is unbiased for \eqref{eq:Qjdef} at every $N_s$, apart from
the multiple of $\bs{I}$ ($N_s$ times the estimation-noise variance) that
\ac{SRS} noise adds, and $\wbar{N}$ appears below only as a dimension.  The
\ac{SRS} cost per \ac{UE} thus grows with $N_s$, each sounding covering $N_s$
antennas with orthogonal pilots (charged in Table~\ref{tab:complexity}), and
$\wh{\bs{Q}}_j$ has generic rank $\min(\wbar{N},N_{\subsf rx})$, the rank used in
Proposition~\ref{prop:subspace}.}

The update time for the matrix estimation, $T_{\subsf LT}$, can be relatively long---on the order of the coherence of the \emph{large-scale propagation parameters}, such as the angles of arrival and path gains, not the phases of the paths.

\revision{%
\begin{remark}[On any factor of $\bs{Q}_i$] \label{rem:Gioptfact}
If $\bs{Q}_i = c\,\bs{B}_i\bs{B}_i\herm$ for any matrix $\bs{B}_i$ and
scalar $c>0$, then
\begin{equation} \label{eq:Gioptfact}
    \bs{G}_i = [\bs{B}_i\herm\bs{Q}^{-1/2}]_r\bs{Q}^{-1/2}
\end{equation}
also maximizes $\wbar{C}_i$. Note that $\sqrt{c}\,\bs{B}_i$ is a square root of
$\bs{Q}_i$, and by Lemma~\ref{lem:capupper} the scaling is irrelevant.  The base station never observes
$\bs{Q}_i$ directly; it observes the $N_{\subsf rx}\times\wbar{N}$ \ac{SRS}
matrix $\wh{\bs{H}}_i$ of \eqref{eq:QHprod}, which is by construction such a
factor with $c = 1/N_{\subsf SRS}$.  Equation \eqref{eq:Gioptfact} therefore
lets the implementation use $\wh{\bs{H}}_i$ as it stands and skip taking the
square root of $\wh{\bs{Q}}_i$ in \eqref{eq:Giopt}---an
$\mc{O}(N_{\subsf rx}^3)$ eigendecomposition per \ac{UE} that would otherwise
dominate the per-\ac{UE} long-term cost---while reducing the singular value
decomposition from $N_{\subsf rx}\times N_{\subsf rx}$ to
$N_{\subsf rx}\times\wbar{N}$.  This is the form actually implemented, and the
one whose cost is reported in Table~\ref{tab:complexity}; it is proved in
Appendix~\ref{sec:proof_projection}.
\end{remark}}

\subsection{Computation of the Matrix Inverse}
\label{sec:matinv}

After computing
$\wh{\bs{Q}}_j$, we need
the inverse 
of the matrix:
\begin{equation} \label{eq:Qhatdef}
    \wh{\bs{Q}} = \bs{I} + \sum_{j=1}^{N_{\subsf UE}} \alpha_j \wh{\bs{Q}}_j,
\end{equation}
which serves as an estimate of \eqref{eq:Qdef}.  As discussed above,
the brute force inversion of this matrix
requires $\mathcal{O}(N_{\subsf rx}^3)$ \acp{FLOP},
which is computationally prohibitive and incompatible with hardware acceleration.
\revision{
\begin{remark}[Reduced-dimension inversion by the Woodbury identity]
\label{rem:woodbury}
Equation \eqref{eq:Qhatdef} is an identity plus a sum of $N_{\subsf UE}$ terms
of rank at most $\wbar{N}$, so its inverse admits a reduced-dimension form as
well.  Stack the $\alpha$-scaled \ac{SRS} matrices of \eqref{eq:Hhatdef},
\begin{equation} \label{eq:Halpha}
    \wh{\bs{H}}_\alpha :=
    \left[\sqrt{\tfrac{\alpha_1}{N_{\subsf SRS}}}\,\wh{\bs{H}}_1,
    \ldots,
    \sqrt{\tfrac{\alpha_{N_{\subsf UE}}}{N_{\subsf SRS}}}\,
    \wh{\bs{H}}_{N_{\subsf UE}}\right]
    \;\in\; \C^{N_{\subsf rx}\times \wbar{N}N_{\subsf UE}},
\end{equation}
so that \eqref{eq:QHprod} and \eqref{eq:Qhatdef} read
$\wh{\bs{Q}} = \bs{I}_{N_{\subsf rx}} +
\wh{\bs{H}}_\alpha\wh{\bs{H}}_\alpha\herm$, and the Woodbury identity gives
\begin{equation} \label{eq:woodbury}
    \wh{\bs{Q}}^{-1}
    = \bs{I}_{N_{\subsf rx}}
    - \wh{\bs{H}}_\alpha
      \left(\bs{I}_{\wbar{N}N_{\subsf UE}}
            + \wh{\bs{H}}_\alpha\herm\wh{\bs{H}}_\alpha\right)^{-1}
      \wh{\bs{H}}_\alpha\herm .
\end{equation}
The $N_{\subsf rx}\times N_{\subsf rx}$ inverse is thereby traded for one of
dimension $\wbar{N}N_{\subsf UE}$, at
$\mc{O}(N_{\subsf rx}(\wbar{N}N_{\subsf UE})^2)$ to form the Gram matrix plus
$\mc{O}(\chi_{\subsf I}(\wbar{N}N_{\subsf UE})^3)$ to invert it---the same
trade the Gram-domain form \eqref{eq:Fgram} makes on the instantaneous side,
with the aggregate \ac{SRS} dimension $\wbar{N}N_{\subsf UE}$ in place of the
joint stream dimension $N_{\subsf UE}N_s$.

The trade is favorable exactly when
$\wbar{N}N_{\subsf UE} < N_{\subsf rx}$, with crossover at
$N_{\subsf UE} = N_{\subsf rx}/\wbar{N}$.  At the parameters of
Table~\ref{tab:parameters} ($\wbar{N} = 8$, $N_{\subsf UE} = 10$) the aggregate
\ac{SRS} dimension is $\wbar{N}N_{\subsf UE} = 80$, so the condition holds
comfortably on the $16\times16$ and $32\times32$ arrays---the crossover
there is at $32$ and $128$ \acp{UE}---and fails only on the $8\times8$ array,
where it is at $8$.  Beyond the crossover the
reduced-dimension inverse is the larger of the two and the identity offers
no saving.  The regime of interest for this paper is therefore the one in which
$\wbar{N}N_{\subsf UE}$, not $N_{\subsf rx}$, is the intrinsic dimension of the
long-term solve.

Equation \eqref{eq:woodbury} yields
$\wh{\bs{Q}}^{-1}$, all that the \ac{CG} route of Sec.~\ref{sec:cgmethod}
uses, through $\bs{A}_i = \wh{\bs{Q}}^{-1}\wh{\bs{H}}_i$.  The inverse
square root of the polynomial route has an exact reduced form too: with the
thin \ac{SVD} $\wh{\bs{H}}_\alpha = \bs{V}\bs{\Sigma}\bs{W}\herm$,
$\wh{\bs{Q}}^{-1/2} = \bs{I} + \bs{V}[(\bs{I}+\bs{\Sigma}^2)^{-1/2}-\bs{I}]\bs{V}\herm$,
at the cost of a $\wbar{N}N_{\subsf UE}$-dimensional eigendecomposition.

The factor
$(\bs{I}+\wh{\bs{H}}_\alpha\herm\wh{\bs{H}}_\alpha)^{-1}$, like that
eigendecomposition, is still a factorization with divisions and sequential data
dependencies, the operation this section sets out to avoid and the reason the
$\mc{O}(N_{\subsf rx}^3)$ inverse above is ruled out.  At
the $N_{\subsf rx} = 1024$ operating point of Table~\ref{tab:complexity},
\eqref{eq:woodbury} is exact and about $50$ and $600$ times cheaper in
long-term arithmetic than the \ac{CG} and polynomial routes.  We nevertheless
use \ac{CG} and the polynomial expansion because both reduce to dense
matrix--matrix products, which map onto the systolic array of
Sec.~\ref{sec:hardware}, and their remaining error is bounded by
Propositions~\ref{prop:subspace} and~\ref{prop:error} rather than set by the
conditioning of a direct solve.  Equation~\eqref{eq:cg_rhs} below exploits the
reduced dimension of \eqref{eq:woodbury} without forming an inverse.
\end{remark}

}
We consider two methods to invert the covariance matrix.

\subsubsection{Conjugate Gradient Method}
\label{sec:cgmethod}
The first method we consider is the \ac{CG} method, a widely used and well-studied iterative method in the \ac{MIMO} \ac{VLSI} detection literature\cite{wu2013approximate, yin2015vlsi}, which computes $\bs{X} \approx \wh{\bs{Q}}^{-1}$.
\revision{$\bs{X}$ cannot simply replace $\wh{\bs{Q}}^{-1/2}$ in
$\bs{G}_i = [\wh{\bs{H}}_i\herm\wh{\bs{Q}}^{-1/2}]_r\wh{\bs{Q}}^{-1/2}$ of
Remark~\ref{rem:Gioptfact}, because $\bs{X}$ approximates
$\wh{\bs{Q}}^{-1}$.  By the invariance noted in
Lemma~\ref{lem:capupper}, $\bs{G}_i$ matters only through its row space, and the
rows of the optimal $\bs{G}_i$ in \eqref{eq:Giopt} span the $r$ principal
eigenvectors of $\wh{\bs{Q}}^{-1}\wh{\bs{Q}}_i$, so any basis of that
eigenspace suffices and $\bs{X}$ alone yields one.  For Hermitian $\bs{X}$
(a finite \ac{CG} iterate is symmetrized, Sec.~\ref{sec:error}), define the
$\wbar{N}\times\wbar{N}$ Hermitian matrix
\begin{equation} \label{eq:Mdef}
    \bs{M}_i \;:=\; \wh{\bs{H}}_i\herm \bs{X} \wh{\bs{H}}_i,
\end{equation}
let $\bs{\Phi}_i \in \C^{\wbar{N}\times r}$ hold its $r$ principal
eigenvectors, and set
\begin{equation} \label{eq:GioptAPP}
    \bs{G}_i \;=\; \bs{\Phi}_i\herm \wh{\bs{H}}_i\herm \bs{X}.
\end{equation}
Writing $\bs{A}_i = \bs{X}\wh{\bs{H}}_i$ and using \eqref{eq:QHprod}, the $j$th
row of $\bs{G}_i$ is $(\bs{A}_i\bs{\varphi}_j)\herm$ and
\begin{equation} \label{eq:Gicanon_exact}
    \bs{X}\wh{\bs{Q}}_i\,(\bs{A}_i\bs{\varphi}_j)
    = \tfrac{1}{N_{\subsf SRS}}\bs{X}\wh{\bs{H}}_i\bs{M}_i\bs{\varphi}_j
    = \tfrac{\lambda_j}{N_{\subsf SRS}}\,(\bs{A}_i\bs{\varphi}_j),
\end{equation}
so its rows are the $r$ principal eigenvectors of
$\bs{X}\wh{\bs{Q}}_i$, and at convergence $\bs{X} = \wh{\bs{Q}}^{-1}$ makes
\eqref{eq:GioptAPP} coincide with the optimum of Lemma~\ref{lem:capupper}
exactly whenever $r\leq\rank(\wh{\bs{Q}}_i)$.  The only error is that of the
solve; Propositions~\ref{prop:subspace} and~\ref{prop:error} bound its effect
on the subspace and on the \ac{SINR}.  The
eigendecomposition required is of $\bs{M}_i$---$8\times8$ in our setting---rather than of an $N_{\subsf rx}\times N_{\subsf rx}$ matrix.

Equation \eqref{eq:GioptAPP} uses $\bs{X}$ only through the
$N_{\subsf rx}\times\wbar{N}$ block $\bs{A}_i$, so there is no reason to form
$\bs{X}$ at all: \ac{CG} can be run on the $\wbar{N}$ right-hand sides of
\begin{equation} \label{eq:cg_rhs}
    \wh{\bs{Q}}\bs{A}_i = \wh{\bs{H}}_i
\end{equation}
instead of on the $N_{\subsf rx}$ columns of the identity.  This changes the
cost from $\mc{O}(tN_{\subsf rx}^3)$ to
$\mc{O}(tN_{\subsf rx}^2\wbar{N}N_{\subsf UE})$ and removes the cubic term from
the long-term budget of Sec.~\ref{sec:complexity}.  At a fixed
$t$ the two are not the same approximation---\ac{CG} applies a Krylov
polynomial whose coefficients depend on the right-hand side---but both
converge to the same limit, and Proposition~\ref{prop:subspace} is stated for
whichever block $\bs{A}_i$ is produced, so it covers either.

Block \ac{CG} therefore runs in one of two regimes, selected by the same
comparison that governs Remark~\ref{rem:woodbury}.

\emph{Regime 1, $\wbar{N}N_{\subsf UE} < N_{\subsf rx}$.}  The \ac{SRS} blocks
are the smaller right-hand side.  The solver is run on \eqref{eq:cg_rhs}, once
per \ac{UE} or on all \acp{UE} jointly, $\wh{\bs{Q}}^{-1}$ is never formed, and
the long-term solve costs
$\mc{O}(tN_{\subsf rx}^2\wbar{N}N_{\subsf UE})$.  This is the regime of the
parameters of Table~\ref{tab:parameters} on the $16\times16$ and $32\times32$
arrays; it is what the implementation does and what
Table~\ref{tab:complexity} charges.

\emph{Regime 2, $\wbar{N}N_{\subsf UE} \geq N_{\subsf rx}$.}  The identity is
the smaller right-hand side.  The solver is then run on
$\wh{\bs{Q}}\bs{X} = \bs{I}$ once, at $\mc{O}(tN_{\subsf rx}^3)$, and every \ac{UE} reads off
$\bs{A}_i = \bs{X}\wh{\bs{H}}_i$ for \eqref{eq:GioptAPP} from that single
$\bs{X}$ at $\mc{O}(N_{\subsf rx}^2\wbar{N})$.  Serving many \acp{UE}, several streams each,
or sounding often enough that the \ac{SRS} blocks outgrow the array puts the
system here; on the arrays evaluated in Sec.~\ref{sec:results} only the
$8\times8$ case does, at $\wbar{N}N_{\subsf UE} = 80$ against
$N_{\subsf rx} = 64$.

The long-term solve therefore costs
$\mc{O}(tN_{\subsf rx}^2\min(\wbar{N}N_{\subsf UE},N_{\subsf rx}))$ in either
case, and the choice is a runtime comparison of two known integers.}
Although the \ac{CG} method theoretically requires $t=N_{\subsf rx}$ iterations to converge to the exact solution, it yields a high-quality approximation in fewer iterations when the system matrix is low rank \cite{wu2013approximate} or the eigenvalues are clustered, as is the case for $\wh{\bs{Q}}$ in
\eqref{eq:Qhatdef} \cite{nocedal2006numerical}.  The
approximation is validated in simulation and analyzed below.

The block \ac{CG} recursion \cite{nocedal2006numerical,o1980block} is the same
in the two regimes; we write it for \eqref{eq:cg_rhs}, the full-inverse regime
replacing $\wh{\bs{H}}_i$ by $\bs{I}$.  \revision{Here $t$ is the iteration
index, $\bs{D}^{(t)}$ the block of search directions, and $\omega_{tj}$,
$\nu_{tj}$ the step size and update coefficient of the $j$th right-hand side.}
From the initial residual
$\bs{D}^{(0)} = \bs{R}^{(0)} = \wh{\bs{H}}_i - \wh{\bs{Q}}\bs{X}^{(0)}$,
\begin{align}
    \bs{X}^{(t+1)} &= \bs{X}^{(t)} + \bs{D}^{(t)} \diag(\bs{\omega}_t),
    \qquad \bs{S}^{(t)} = \wh{\bs{Q}}\bs{D}^{(t)}, \label{eq:CG_K}\\
    \bs{R}^{(t+1)} &= \wh{\bs{H}}_i - \wh{\bs{Q}}\bs{X}^{(t+1)}, \nonumber\\
    \bs{D}^{(t+1)} &= \bs{R}^{(t+1)} + \bs{D}^{(t)} \diag(\bs{\nu}^{(t)}),
    \label{eq:CG_P}
\end{align}
with the step size and update coefficient
\begin{equation} \label{eq:CG_omega}
    \omega_{tj} = \frac{\|\bs{r}^{(t)}_j\|^2}{(\bs{d}^{(t)}_j)\herm \bs{s}^{(t)}_j},
    \qquad
    \nu_{tj} = \frac{\|\bs{r}^{(t+1)}_j\|^2}{\|\bs{r}^{(t)}_j\|^2},
\end{equation}
where $\bs{r}^{(t)}_j$, $\bs{d}^{(t)}_j$, $\bs{s}^{(t)}_j$ denote the $j$th
columns of $\bs{R}^{(t)}$, $\bs{D}^{(t)}$ and $\bs{S}^{(t)}$.
The benefit of \ac{CG} is that the bulk of the operations are in the matrix multiplication step
\eqref{eq:CG_K}, which can be realized
efficiently in hardware via a systolic array---an architecture used in many of the \ac{VLSI} 
implementations described above.

\subsubsection{Polynomial Approximation Method}
\label{sec:polymethod}

\ac{CG} relies on sequential vector operations that typically require a \ac{DSP}
or custom logic.  To keep the computation on a systolic array, optimized for dense matrix--matrix products, we also consider the \revision{order-$d$}
polynomial approximation
\begin{equation} \label{eq:ppoly}
    \bs{P}(\bs{\beta}) := \sum_{\ell=0}^{d-1} \beta_\ell \wh{\bs{Q}}^\ell
\end{equation}
with the coefficients $\bs{\beta}$ chosen so that
$\bs{P}(\bs{\beta}) \approx \wh{\bs{Q}}^{-1/2}$, by minimizing the error
\begin{equation} \label{eq:Jbeta1}
    \mc{J}(\bs{\beta}) := \| \bs{P}(\bs{\beta})\wh{\bs{Q}}\bs{P}(\bs{\beta}) -\bs{I}\|,
\end{equation}
where $\|\cdot\|$ is \revision{the spectral norm}; this objective is justified in
Sec.~\ref{sec:error}.  By the spectral mapping theorem \cite{horn2012matrix} and
$\bs{I} \preceq \wh{\bs{Q}} \preceq \lambda_{\max}\bs{I}$, minimizing \eqref{eq:Jbeta1} reduces to a
Chebyshev-type problem on $[1,\lambda_{\max}]$ whose data-independent solution is computed
offline once by the Remez exchange algorithm
\cite{powell1981approximation}, leaving only the
polynomial multiplication \eqref{eq:ppoly} online.  Given $\bs{P}$, we
compute $\bs{G}_i$ from Lemma~\ref{lem:capupper} with $\bs{Q}^{-1/2}$ replaced
by $\bs{P}$ and $\bs{Q}_i$ by $\wh{\bs{Q}}_i$:
\begin{equation} \label{eq:Giapprox}
    \bs{G}_i \revision{=} [\wh{\bs{H}}_i\herm \revision{\wh{\bs{Q}}}^{-1/2} ]_r \revision{\wh{\bs{Q}}}^{-1/2}
    \revision{\approx} [\wh{\bs{H}}_i\herm \bs{P} ]_r \bs{P},
\end{equation}
where, again, we have used \eqref{eq:QHprod}.

\revision{%
\subsection{Signaling and Overhead in 5G NR}
\label{sec:signalling}

We now map the $N_{\subsf SRS}$ \ac{SRS} snapshots and the in-band \ac{DM-RS}
onto actual \ac{NR} resources.  An \ac{SRS} resource occupies every
$K_{\subsf TC}$th subcarrier of one or more \ac{OFDM} symbols, with \acp{UE}
separated within a comb offset by cyclic shifts of the base sequence
\cite[Sec.~6.4.1.4]{3gpp38211}, so one \ac{SRS} symbol carries
$K_{\subsf TC}n_{\subsf cs}$ orthogonal ports---$48$ for the comb-$4$ and
$12$ cyclic shifts we assume.  Sounding $N_{\subsf UE}N_s$ antenna
ports once per $T_{\subsf LT}$ costs
\begin{equation} \label{eq:srssym}
    N^{\subsf SRS}_{\subsf sym}
    = \left\lceil \frac{N_{\subsf UE}N_s}{K_{\subsf TC}n_{\subsf cs}}\right\rceil
    N^{\subsf SRS}_{\subsf symb}
\end{equation}
\ac{OFDM} symbols, with $N^{\subsf SRS}_{\subsf symb}$ the repetition factor of
the resource.  At the parameters of Table~\ref{tab:parameters}
($N_{\subsf UE}N_s = 10$, $T_{\subsf LT}=\SI{10}{ms}$, which is $1120$ symbols
at \SI{120}{kHz} \ac{SCS}) that is a single \ac{SRS} symbol per
long-term period: $0.089\%$ of the grid if the scheduler blanks the
whole symbol, and $0.022\%$ if it rate-matches \ac{PUSCH} around the
comb.  \revision{Equation~\eqref{eq:srssym} counts symbols per $T_{\subsf LT}$
and is numerology-independent; only the grid fraction
is not.  At the \SI{30}{kHz} spacing of \ac{FR1}, where $T_{\subsf LT}$ is
$280$ symbols, the same single \ac{SRS} symbol is $0.36\%$ of the
grid, and a $1\%$ \ac{SRS} budget still supports $96$ antenna ports per
$T_{\subsf LT}$.}  The \ac{SRS} pilot processing gain $G_{\subsf SRS}$ that these resources deliver is derived
from the measured coherence bandwidth in Sec.~\ref{sec:results}.

\ac{DM-RS} is a different matter.  Both receivers demodulate the same
$N_{\subsf UE}N_s$ layers and therefore need the same orthogonal \ac{DM-RS}
ports at the same density, so the \ac{DM-RS} overhead is the same for both
receivers, and the total pilot cost of \ac{LTBF} is higher by the
\ac{SRS} term of \eqref{eq:srssym}: the reduction \ac{LTBF} achieves is in the
coefficient refresh rate, not the pilot budget, and Table~\ref{tab:pilot}
reports both.  Ten layers require the double-symbol type-2 configuration, \revision{up to
Release~17} the only one reaching $12$ orthogonal ports
\cite[Sec.~6.4.1.1]{3gpp38211} \revision{(Release~18 allows $24$)}; it spends
$2$ of the $14$ symbols of a slot, or $14.3\%$, against which
the $0.022\%$ of \ac{SRS} is negligible.  From the
same pilots the reduced domain gains estimation accuracy, since an
$r$-dimensional effective channel replaces an $N_{\subsf rx}$-dimensional one;
on the already minimal, quantized \ac{DM-RS} grid this adds margin but does not
save \acp{RE}.

The port ceiling also limits the arithmetic advantage of
Sec.~\ref{sec:complexity}: $N_{\subsf UE}N_s \leq 12$ \revision{($24$ in Release~18)} is the largest group a
standard-compliant deployment can demodulate without pilot reuse, so the
crossover with the Gram-domain baseline, at $N_{\subsf UE}N_s\approx15$ with the
\ac{CG} solve and $50$ with the polynomial, lies beyond one \ac{NR}
\ac{MU-MIMO} group up to Release~17, and only the former lies within the $24$
ports of Release~18.

\begin{table}[t]
\centering
\caption{\revision{Pilot cost and coefficient rate of the two receivers against $N_{\subsf UE}N_s$, at the parameters of Table~\ref{tab:parameters}. $\dagger$: beyond the $12$ orthogonal \ac{DM-RS} ports of TS~38.211 up to Release~17 (Release~18 allows $24$).}}
\scriptsize
\label{tab:pilot}
\setlength{\tabcolsep}{2pt}
\begin{tabular}{@{}c|cc|cc|cc@{}}
\toprule
 & \multicolumn{2}{c|}{\textbf{\acs{SRS}} (\acs{LTBF} only)} & \multicolumn{2}{c|}{\textbf{\acs{DM-RS}} (both)} & \multicolumn{2}{c}{\textbf{Coeff. rate} (s$^{-1}$)} \\
$N_{\subsf UE}N_s$ & \makecell{sym.\\per $T_{\subsf LT}$} & \% grid & \makecell{\% grid\\$T_{\subsf coh}\!=\!28$} & \makecell{\% grid\\$T_{\subsf coh}\!=\!7$} & inst. & \acs{LTBF} \\
\midrule
4 & 1 & 0.022 & 14.3 & 28.6 & $1.1\!\times\!10^{9}$ & $8.2\!\times\!10^{5}$ \\
10 & 1 & 0.022 & 14.3 & 28.6 & $2.7\!\times\!10^{9}$ & $2.0\!\times\!10^{6}$ \\
12 & 1 & 0.022 & 14.3 & 28.6 & $3.2\!\times\!10^{9}$ & $2.5\!\times\!10^{6}$ \\
24$^\dagger$ & 1 & 0.045 & 28.6 & 57.1 & $6.5\!\times\!10^{9}$ & $4.9\!\times\!10^{6}$ \\
48$^\dagger$ & 1 & 0.089 & 57.1 & 100.0 & $1.3\!\times\!10^{10}$ & $9.8\!\times\!10^{6}$ \\
\bottomrule
\end{tabular}
\end{table}

}

\begin{table*}[t]
\centering
\caption{Cost of one carrier of uplink processing, in \acp{CMAC}; one \ac{CMAC} is $8$ real operations. $N_s=1$ and $r=1$, other parameters of Table~\ref{tab:parameters}, hardware of Sec.~\ref{sec:hardware}.}
\scriptsize
\label{tab:complexity}
\renewcommand{\arraystretch}{1.0}
\setlength{\tabcolsep}{3pt}
\begin{tabular}{@{}p{1.5cm}|p{4.3cm}@{}||>{\centering\arraybackslash}p{2.4cm}|>{\centering\arraybackslash}p{2.4cm}|>{\centering\arraybackslash}p{2.4cm}|>{\centering\arraybackslash}p{2.4cm}|>{\centering\arraybackslash}p{1.2cm}@{}}
\toprule
 \textbf{Budget} & \textbf{Operation} & \textbf{\makecell{Inst. \acs{MMSE}\\full-dim.~\eqref{eq:Finstant}}} & \textbf{\makecell{Inst. \acs{MMSE}\\Gram~\eqref{eq:Fgram}\\\cite{jeon2019decentralized}}} & \textbf{\makecell{\acs{LTBF}\\(order $d$)}} & \textbf{\makecell{\acs{LTBF}\\(\acs{CG}, $t$ iterations,\\right-hand-side\\solve)}} & \textbf{Rate} \\
\midrule
\multirow{4}{1.5cm}{\raggedright\textbf{\acs{CSI} acquisition}}
& \textbf{Full-dimensional channel estimation} & $N_{\subsf rx} N_{\subsf UE} N_{s} N_{\subsf DM\text{-}RS}$ & $N_{\subsf rx} N_{\subsf UE} N_{s} N_{\subsf DM\text{-}RS}$ & -- & -- & $1/T_{\subsf coh}$ \\
& \textbf{\acs{SRS} channel estimation} & -- & -- & $N_{\subsf rx} \wbar{N} N_{\subsf UE}$ & $N_{\subsf rx} \wbar{N} N_{\subsf UE}$ & $1/T_{\subsf LT}$ \\
& \textbf{Reduced-dimension channel estimation} & -- & -- & $r N_{s} N_{\subsf DM\text{-}RS} N_{\subsf UE}$ & $r N_{s} N_{\subsf DM\text{-}RS} N_{\subsf UE}$ & $1/T_{\subsf coh}$ \\
\cmidrule(lr){2-7}
& \textbf{GFLOP/s per carrier} & 173 & 173 & 0.234 & 0.234 & -- \\
\midrule
\multirow{10}{1.5cm}{\raggedright\textbf{Beamformer computation}}
& \textbf{Covariance $\bs{Q} = \bs{I} + \bs{H}\bs{H}\herm$} & $N_{\subsf rx}^{2} N_{\subsf UE} N_{s}$ & -- & -- & -- & $1/T_{\subsf coh}$ \\
& \textbf{Inversion $\bs{Q}^{-1}$} & $\chi_{\subsf I} N_{\subsf rx}^{3}$ & -- & -- & -- & $1/T_{\subsf coh}$ \\
& \textbf{Combiner $\bs{F} = \bs{Q}^{-1}\bs{H}$} & $N_{\subsf rx}^{2} N_{\subsf UE} N_{s}$ & -- & -- & -- & $1/T_{\subsf coh}$ \\
& \textbf{Gram matrix $\bs{H}\herm\bs{H}$} & -- & $N_{\subsf rx} (N_{\subsf UE} N_{s})^{2}$ & -- & -- & $1/T_{\subsf coh}$ \\
& \textbf{Inversion $(\bs{I} + \bs{H}\herm\bs{H})^{-1}$} & -- & $\chi_{\subsf I} (N_{\subsf UE} N_{s})^{3}$ & -- & -- & $1/T_{\subsf coh}$ \\
& \textbf{Estimating $\bs{Q}$~\eqref{eq:Qdef}} & -- & -- & $\wbar{N} N_{\subsf rx}^{2} N_{\subsf UE}$ & $\wbar{N} N_{\subsf rx}^{2} N_{\subsf UE}$ & $1/T_{\subsf LT}$ \\
& \textbf{Computing $\bs{Q}^{-1/2}$} & -- & -- & $(d-1) N_{\subsf rx}^{3}$ & $t N_{\subsf rx}^{2} \wbar{N} N_{\subsf UE}$ & $1/T_{\subsf LT}$ \\
& \textbf{Computing $\bs{G}_i$~\eqref{eq:Giopt}} & -- & -- & $\big[(\wbar{N}+r) N_{\subsf rx}^{2} + \chi_{\subsf EVD} \wbar{N}^{3}\big] N_{\subsf UE}$ & $\big[(\wbar{N}+r) \wbar{N} N_{\subsf rx} + \chi_{\subsf EVD} \wbar{N}^{3}\big] N_{\subsf UE}$ & $1/T_{\subsf LT}$ \\
& \textbf{Reduced-dimension inversion} & -- & -- & $\chi_{\subsf I} r^{3} N_{\subsf UE}$ & $\chi_{\subsf I} r^{3} N_{\subsf UE}$ & $1/T_{\subsf coh}$ \\
\cmidrule(lr){2-7}
& \textbf{GFLOP/s per carrier} & $4.58\!\times\!10^{6}$ & 220 & 7874 & 605 & -- \\
\midrule
\multirow{5}{1.5cm}{\raggedright\textbf{Per-\acs{RE} data path}}
& \textbf{Equalization} & $N_{\subsf rx} N_{\subsf UE} N_{s}$ & -- & $r N_{s} N_{\subsf UE}$ & $r N_{s} N_{\subsf UE}$ & $1/T_{\subsf RE}$ \\
& \textbf{\acs{MRC} reduction $\bs{H}\herm\bs{y}$} & -- & $N_{\subsf rx} N_{\subsf UE} N_{s}$ & -- & -- & $1/T_{\subsf RE}$ \\
& \textbf{Reduced-dimension equalization} & -- & $(N_{\subsf UE} N_{s})^{2}$ & -- & -- & $1/T_{\subsf RE}$ \\
& \textbf{Low-rank projection~\eqref{eq:zproj}} & -- & -- & $r N_{\subsf rx} N_{\subsf UE}$ & $r N_{\subsf rx} N_{\subsf UE}$ & $1/T_{\subsf RE}$ \\
\cmidrule(lr){2-7}
& \textbf{GFLOP/s per carrier} & 7267 & 7338 & 7274 & 7274 & -- \\
\midrule
\multirow{3}{1.5cm}{\raggedright\textbf{Total}} & \textbf{GFLOP/s per carrier} & $4.59\!\times\!10^{6}$ & 7731 & $1.51\!\times\!10^{4}$ & 7879 & -- \\
& \textbf{Reference-array occupancy} & 3564 & 3.02 & 5.92 & 3.08 & -- \\
& \textbf{Coefficient refresh rate (s$^{-1}$)} & $2.70\!\times\!10^{9}$ & $2.70\!\times\!10^{9}$ & $1.02\!\times\!10^{6}$ & $1.02\!\times\!10^{6}$ & -- \\
\bottomrule
\end{tabular}
\end{table*}

\section{Complexity Analysis and Hardware Implementation}
\label{sec:complexity}

\subsection{Computational Complexity Scaling}
\label{sec:cxscaling}
To assess scalability, we estimate the computational complexity in terms of
dominant scaling terms and update frequency.  The load is decoupled into
\revision{three} distinct time scales: long-term centralized processing once per
$T_{\subsf LT}$, per-coherence-block processing at $T_{\subsf coh}$, and
per-resource-element processing at $T_{\subsf RE}$.

\revision{%
Table~\ref{tab:complexity} reports the cost of one carrier of uplink processing
for each scheme, decomposed into the three budgets above, at $r=1$, which in
the rural rank sweep with a noiseless \ac{SRS} estimate loses only \SI{0.04}{dB} against the $r=2$ used
elsewhere (Sec.~\ref{sec:coverror}).

\emph{(i) Against the full-dimensional form \eqref{eq:Finstant}}, the proposed
method reduces the total arithmetic load by $3.0$--$5.8\times10^{2}$ at this
operating point ($N_{\subsf rx}=1024$, $N_{\subsf UE}N_s = 10$,
$r=1$), and the beamformer-computation budget by over
$5\times10^{2}$.

\emph{(ii) Against the Gram-domain form \eqref{eq:Fgram}}, the relevant
baseline, the proposed method increases the total operation count at this
operating point, by $2\%$ with the
\ac{CG} solve and $2.0\times$ with the polynomial.  The per-\ac{RE} rows
nearly coincide---the projection costs
$rN_{\subsf rx}N_{\subsf UE}$ operations per \ac{RE} against
$N_{\subsf rx}N_{\subsf UE}N_s$ for the \ac{MRC} reduction, a ratio $r/N_s$
of $1$ here and $2$ at $r=2$---so the excess is in the long-term beamformer
budget.  The same model, evaluated as $N_{\subsf UE}$ increases at $N_s=1$ and $r=1$, crosses over
at $N_{\subsf UE}N_s \approx 15$ with the \ac{CG} solve and $50$ with the
polynomial, where the
baseline's Gram matrix, inverse and per-\ac{RE} terms, superlinear in
$N_{\subsf UE}N_s$, overtake the proposed method.

\emph{(iii) In \ac{CSI} acquisition}, the Gram-domain receiver must produce a fresh
$N_{\subsf rx}\times N_{\subsf UE}N_s$ combining matrix once per coherence time
and once per coherence bandwidth; the proposed method produces one
frequency-flat $r\times N_{\subsf rx}$ projection per \ac{UE} once per
$T_{\subsf LT}$.  At this operating point, that is
$2.7\times10^{9}$ against $1.02\times10^{6}$ complex coefficients per second, a
reduction of $2.6\times10^{3}$ ($1.3\times10^{3}$ at the $r=2$ of
Table~\ref{tab:pilot}), together with a $7\times10^{2}$ reduction in
channel-estimation arithmetic.  This reduction of the coefficient rate, from
$\mc{O}(N_{\subsf rx}N_s/T_{\subsf coh})$ to $\mc{O}(rN_{\subsf rx}/T_{\subsf LT})$
per \ac{UE}, is the main gain of the method; it also makes the real-time path
weight-stationary and decoupled across \acp{UE}.}

The complexity of each operation is expressed in terms of system parameters, the values of which are listed in Table~\ref{tab:parameters}.

\revision{%
Two conventions make Table~\ref{tab:complexity} readable.  All counts are
complex multiply-accumulate operations, each eight real operations, so the
\ac{FLOP} figures are eight times the operation counts.  An operation labeled $1/T_{\subsf coh}$ is
executed once per coherence block---once per $T_{\subsf coh}$ symbols
and once per coherence bandwidth---whereas one labeled
$1/T_{\subsf LT}$ is executed once per carrier per long-term period, the
long-term statistics being frequency-flat.  That distinction is the single
largest term in the comparison and the reason the \ac{CSI}-acquisition budget
separates the two schemes by three orders of magnitude.}

Relative to their update rates, the dominant terms therefore fall as follows.
\revision{The full-dimensional receiver \eqref{eq:Finstant} incurs an
$\mc{O}(N_{\subsf rx}^3)$ term once per $T_{\subsf coh}$, where it is the
real-time bottleneck; the Gram-domain form \eqref{eq:Fgram} replaces that term
by $\mc{O}(N_{\subsf rx}(N_{\subsf UE}N_s)^2 +
\chi_{\subsf I}(N_{\subsf UE}N_s)^3)$ at the same rate, still on the real-time
path and still coupled across \acp{UE}.  The proposed method performs its
beamformer computation once per $T_{\subsf LT}$ instead, in a centralized
long-term update whose amortized per-symbol cost is negligible; with
\eqref{eq:cg_rhs} that update carries no cubic term at all, and the polynomial
route's $\mc{O}(dN_{\subsf rx}^3)$ is charged at $1/T_{\subsf LT}$ rather than
$1/T_{\subsf coh}$.}
What is left on the real-time path is
\revision{$\mathcal{O}(r N_{\subsf rx})$ per \ac{UE} per resource element and
$\mathcal{O}(r^3)$ per \ac{UE} per coherence block}, against an
$\mathcal{O}(r N_{\subsf rx})$ footprint: linear in $N_{\subsf rx}$ and
decoupled across users, hence parallelizable across hardware cores as
$N_{\subsf UE}$ grows.

\revision{%
\subsection{Reference Hardware and Accounting Assumptions}
\label{sec:hardware}

Operation counts are converted to time with the following hardware model.  The
reference platform is $N_{\subsf arr} = 20$ instances of the reconfigurable spatial array of our
prior work~\cite{rasteh2025spatial}, synthesized in a SAED \SI{32}{nm} \ac{LVT}
node at \SI{1}{GHz}.  Each array has 16 complex multiply-accumulate units per
cycle and a 16-bit fixed-point datapath, occupies \SI{1.014}{mm^2}, and draws
\SI{193}{mW}.  Its measured throughput is \SI{127.98}{GOPS} at $99.98\%$
utilization on dense complex matrix
multiplication~\cite[Table II]{rasteh2025spatial}.  The platform therefore
sustains $3.20\times10^{11}$ complex multiply-accumulates per second, i.e.,
\SI{2.56}{TOPS} in real operations, at \SI{20.3}{mm^2} and \SI{3.86}{W}.
Kernels that are not dense matrix products are derated to half that
utilization, consistent with reported systolic LU and QR
implementations~\cite{tian2024flud,wu2013approximate}; the
matrix--matrix figure applies because the dominant kernels of
both the proposed method and the baselines have high data reuse.  Off-chip
traffic and memory-bandwidth stalls beyond those reflected in the measured
utilization are not counted, so the occupancy row of
Table~\ref{tab:complexity} is a lower bound.  It is a ratio, not a latency: a
value above unity means that a 20-array platform is oversubscribed.
Because every scheme is evaluated on the same platform with the same derating,
only the ratios between columns are claimed, \revision{and even these
depend on the kernel mix (Sec.~\ref{sec:runtime})}; the absolute values scale
inversely with $N_{\subsf arr}$.

\subsection{Measured Runtime}
\label{sec:runtime}

We also measured wall-clock times for the dominant kernels and for one coherence
block of each scheme, at $r=2$, so that the analytic model can be checked against an
implementation.  These are general-purpose-processor numbers---single-threaded
\texttt{complex64} \ac{BLAS} on one core of an AMD EPYC 7502, median of seven
runs after a discarded warm-up---and they are not comparable to the
spatial-array projections of Table~\ref{tab:complexity}; they are reported to
give the practical implementation cost of the proposed method alongside the
baselines, and because the relative kernel costs are a useful check on the
operation counts.

The measurements confirm the analytic picture.  One coherence block of the
full-dimensional receiver \eqref{eq:Finstant} takes about \SI{91}{ms}, against
\SI{0.71}{ms} for the Gram-domain form \eqref{eq:Fgram} and \SI{0.88}{ms} for
the \ac{LTBF} real-time path: a reduction of two orders of magnitude against the
full-dimensional form, and an increase of about $1.2\times$ against the
Gram-domain form, below the $r/N_s=2$ ratio of their operation counts because
the wider projection sustains a higher throughput on this processor.  The long-term
update, in the form \eqref{eq:cg_rhs} that Table~\ref{tab:complexity} charges, takes
about \SI{93}{ms}: on one core it overruns $T_{\subsf LT}=\SI{10}{ms}$, so it
must run in parallel across the per-\ac{UE} solves or over a longer
$T_{\subsf LT}$, and amortized it adds about \SI{35}{\micro\second}, $4\%$, per
coherence block.  The order-$10$ polynomial whitener sustains \SI{97}{GFLOP/s} against
\SI{58}{GFLOP/s} for the equivalent dense inverse.
}

\section{Error Analysis}
\label{sec:error}
\revision{%
The projection actually computed differs from the one
Lemma~\ref{lem:capupper} prescribes in two ways: the covariances are estimated,
which Sec.~\ref{sec:coverror} measures, and the linear system is solved only
approximately, which Proposition~\ref{prop:subspace} bounds.
Proposition~\ref{prop:error} quantifies the effect of the resulting subspace
error on the \ac{SINR}, under the assumptions of Sec.~\ref{sec:assumptions}.

\subsection{Standing Assumptions}
\label{sec:assumptions}

\begin{assumption}[Scope of the \ac{SINR} analysis] \label{as:sinr}
The results of Sec.~\ref{sec:sinrloss} are stated under the following assumptions.
\begin{enumerate}[label=(A\arabic*)]
\item \emph{Rank-one channel, single stream.}  $N_s = 1$ and, over the
long-term period, $\bs{h}_i[n,k] = c_i[n,k]\bs{u}_i$ with $\bs{u}_i$ fixed and
$c_i[n,k]$ zero mean and varying with the small-scale fading, so that
$\bs{Q}_i = g_i\bs{u}_i\bs{u}_i\herm$ with $g_i = \Exp|c_i[n,k]|^2$.
\item \emph{Exact covariance.}  $\wh{\bs{Q}}_i = \bs{Q}_i$ for all $i$, and
hence $\wh{\bs{Q}} = \bs{Q}$.
\end{enumerate}
\end{assumption}

\emph{(A1)} is the case produced by a single dominant path per \ac{UE}, and we
take $r=1$ in Sec.~\ref{sec:sinrloss}.  \emph{(A2)} isolates the solver error from the estimation error and is not innocuous, $\wh{\bs{Q}}_i$ being built from
$\wbar{N}$ noisy \ac{SRS} columns; no result below bounds that error, so we
measure it instead in Sec.~\ref{sec:coverror}.
Proposition~\ref{prop:subspace} needs none of Assumption~\ref{as:sinr}.

Assumption (A1) makes $r>1$ degenerate, $[\cdot]_r$ returning
one meaningful direction and $r-1$ arbitrary ones, and the evaluated scenario is
of that kind (Sec.~\ref{sec:rank}), so $r>1$ captures no further propagation
paths there.  Under (A1)--(A2) the extra rows leave $\gamma_i$ unchanged: in whitened
coordinates they carry no signal, and their interference-plus-noise is
uncorrelated with that of the signal row.  They matter only through model
mismatch: \ac{SRS} noise lifts the nominally zero eigenvalues of
$\wh{\bs{Q}}_i$ to a measured effective rank of $4.8$ at $\wbar{N}=8$ and
$G_{\subsf SRS}=\SI{0}{dB}$, and $1.5$ at the nominal \SI{10}{dB}
(Sec.~\ref{sec:coverror}), and the channel evolves over $T_{\subsf LT}$, so the
extra rows observe interference that the long-term statistics do not capture.
Since the subspaces are nested, increasing $r$ cannot decrease $\gamma_i$.

\subsection{Accuracy of the Computed Projection}
\label{sec:subspace}

Write $\bs{\Psi}_i := \wh{\bs{Q}}^{-1/2}\wh{\bs{H}}_i$ for the whitened \ac{SRS}
block, so that $\wh{\bs{Q}}_i\wtn := \wh{\bs{Q}}^{-\subsf H/2}\wh{\bs{Q}}_i\wh{\bs{Q}}^{-1/2} =
\bs{\Psi}_i\bs{\Psi}_i\herm/N_{\subsf SRS}$ is the estimate of the matrix of
Corollary~\ref{cor:sinr}.  Let $\mu_1 \geq \mu_2 \geq \cdots \geq 0$
be its eigenvalues and $\delta_r := \mu_r - \mu_{r+1}$ the spectral gap at the
projection rank.  Principal angles are measured in the
whitened coordinates $\wh{\bs{Q}}^{1/2}\bs{w}$, in which
the Ritz values---and hence, by Corollary~\ref{cor:sinr}, both objectives---are defined.

\begin{proposition} \label{prop:subspace}
Let $\bs{A}_i \in \C^{N_{\subsf rx}\times\wbar{N}}$ be any computed
approximation of $\wh{\bs{Q}}^{-1}\wh{\bs{H}}_i$, and let
\begin{equation} \label{eq:edef}
    e := \frac{\|\wh{\bs{Q}}^{1/2}\bs{A}_i - \bs{\Psi}_i\|}{\|\bs{\Psi}_i\|}
\end{equation}
be its relative error in the whitened coordinates.  Suppose $\delta_r > 0$ and
$2e\mu_1 \leq \delta_r$.  Then the row space of the projection
\eqref{eq:GioptAPP} computed from $\bs{A}_i$, with $\bs{\Phi}_i$ taken from the
Hermitian part of $\wh{\bs{H}}_i\herm\bs{A}_i$, and that of the maximizer
\eqref{eq:Giopt} at the estimates $\wh{\bs{Q}}_i$, $\wh{\bs{Q}}$ satisfy
\begin{equation} \label{eq:sintheta}
    \sin\Theta \;\leq\; \frac{2\pi\, e\, \mu_1}{\delta_r},
\end{equation}
where $\sin\Theta$ denotes the sine of the largest principal angle between
them.
\end{proposition}
\begin{proof} See Appendix~\ref{sec:proof_subspace}.
\end{proof}

The bound holds for any $\wh{\bs{Q}}_i$, of any rank and from any number of
streams, and for any way of producing $\bs{A}_i$, so it covers both
the full inverse and the right-hand-side solve \eqref{eq:cg_rhs}.  It goes to
zero linearly in $e$, so the computed projection converges to
Lemma~\ref{lem:capupper}'s maximizer as the solver converges, and it degrades
only through $\mu_1/\delta_r$: a rank chosen inside a cluster of nearly equal
Ritz values identifies no well-defined subspace, while a rank chosen at a clear
spectral gap is stable.

\subsection{\acs{SINR} Loss of the Approximate Inverse}
\label{sec:sinrloss}
}
Recall that \revision{the proposed solution of Sec.~\ref{sec:soln} seeks}
a matrix $\bs{P} \approx \bs{Q}^{-1/2}$%
\revision{, or a matrix $\bs{X}\approx\bs{Q}^{-1}$}.
\revision{%
We analyze the effect of this approximation under Assumption~\ref{as:sinr}.}
We compare three spatial projections.  The \emph{ideal} projection uses the
true inverse square root $\bs{Q}^{-1/2}$,
\begin{equation} \label{eq:Giideal}
    \bs{G}_i^0 = [\bs{Q}_i^{1/2}\bs{Q}^{-1/2}]_{\revision{1}}\bs{Q}^{-1/2},
    \quad
    \bs{z}^0_i = \bs{G}^0_i\bs{y} = \bs{G}^0_i( \bs{H}_i\bs{x}_i + \bs{d}_i),
\end{equation}
the \emph{approximate \ac{CG}} projection of \eqref{eq:GioptAPP} uses
$\bs{X} \approx \bs{Q}^{-1}$,
\begin{equation} \label{eq:Giapprox_cg}
    \bs{G}_i = \revision{\bs{u}_i\herm\bs{X}},
    \qquad
    \bs{z}_i = \bs{G}_i\bs{y} = \bs{G}_i( \bs{H}_i\bs{x}_i + \bs{d}_i),
\end{equation}
\revision{and} the \emph{approximate
polynomial} projection uses $\bs{P} \approx \bs{Q}^{-1/2}$, giving
$\bs{G}_i = [\bs{Q}_i^{1/2}\bs{P}]_{\revision{1}}\bs{P}$, with $\bs{z}_i$ as in \eqref{eq:Giapprox_cg}.  Our main results characterize
the \ac{SINR} loss between the ideal and approximate projections.
\revision{%
Both branches supply a Hermitian $\bs{Y}$---$\bs{Y} = \bs{P}^2$ for the
polynomial filter and $\bs{Y} = (\bs{X}+\bs{X}\herm)/2$ for \ac{CG}, the
symmetrization being both what the implementation forms and what
Appendix~\ref{sec:proof_error} requires---and the premise below bounds how far
$\bs{Y}$ is from $\bs{Q}^{-1}$; $\gamma_i^0$ and $\gamma_i$ denote the
\ac{SINR} \eqref{eq:sinr_proj} under the ideal and the approximate projection.}

\begin{proposition} \label{prop:error}
\revision{Let Assumption~\ref{as:sinr} hold and suppose that, for some
$\epsilon \in (0,1)$, the approximate projection is computed either via \ac{CG}
with the symmetrized iterate $\bs{Y} = (\bs{X}+\bs{X}\herm)/2$, or via a
polynomial spatial filter $\bs{P}$, satisfying respectively}
\begin{equation} \label{eq:cg_premise}
    \| \bs{Q}\bs{Y} - \bs{I}\| < \epsilon,
    \qquad
    \| \bs{P} \bs{Q} \bs{P} - \bs{I} \| \leq \epsilon .
\end{equation}
\revision{%
Then, in any resource element $(n,k)$,
\begin{equation}
    \gamma_i
    \geq
    \frac{\gamma_i^0 (1-\epsilon)^2}{ (1+\epsilon)^2 + 4\epsilon
    \Exp( \gamma_i^0 ) } ,
    \label{eq:sinr_error}
\end{equation}
the expectation being over the small-scale fading $c_i[n,k]$.}
\end{proposition}
\begin{proof} See Appendix~\ref{sec:proof_error}.
\end{proof}

\revision{%
High-\ac{SINR} \acp{UE} are the most sensitive to inverse-approximation error, which
justifies the objective $\mc{J}(\bs{\beta})$ in \eqref{eq:Jbeta1}.  For
$r\geq\rank(\bs{Q}_i)$ the bound extends to $\bs{Q}_i$ of any rank and to any
$N_s$, as a Loewner-order bound on $\wt{\bs{H}}_i\herm\wt{\bs{R}}_i^{-1}\wt{\bs{H}}_i$
in which $\Exp(\gamma_i^0)$ is replaced by the mean \ac{SINR} of the strongest
eigenmode of $\bs{Q}_i\wtn$; we omit the proof.}

\section{Performance Evaluation via Ray-Tracing Simulations} \label{sec:results}

\begin{table}[t]
    \centering
    \scriptsize
    \setlength{\tabcolsep}{4pt}
    \caption{Summary of the parameters employed in simulations.}
    \begin{tabular}{c||c||c}
        \hline
        \textbf{Parameter} & \textbf{Description} & \textbf{Value} \\
        \hline
        $f_{c}$ & Carrier frequency  & \SI{3.5}{GHz} \\
        $N_{\subsf rx}$ & Number of \acs{BS} RX antennas & $\scriptstyle 8 \times 8$--$\scriptstyle 32 \times 32$ \\
        $r$ & Rank of \acs{LTBF} projection matrix & 2 \\
        $N_{\subsf SRS}$ & Number of \acs{SRS} \revision{snapshots} per \acs{UE} & 8 \\
        \revision{$N_s$} & \revision{Streams per \acs{UE}} & \revision{1} \\
        \revision{$\wbar{N}$} & \revision{\acs{SRS} columns per \acs{UE}, $N_sN_{\subsf SRS}$} & \revision{8} \\
        \revision{$G_{\subsf SRS}$} & \revision{\makecell{\acs{SRS} pilot processing gain: how far the \\ estimation-error variance sits below \\ the data-domain noise floor}} & \revision{\SI{10}{dB}} \\
        $d$ & Order of polynomial approximations & 2--20 \\
        \revision{$t$} & Number of iterations in the \acs{CG} method & 3--10 \\
        $\chi_{\subsf EVD}$ & \makecell{Coefficient of matrix eigenvalue \\ decomposition complexity} & 10/3 \\
        $\chi_{\subsf I}$ & Coefficient of matrix inversion complexity & 2 \\
        $N_{\subsf UE}$ & Number of users per sector & 10 \\
        $T_{\subsf LT}$ & Long-term estimation period & \makecell{\SI{10}{ms}\\\revision{($1120$ symbols)}} \\
        \revision{$T_{\subsf coh}$} & \revision{Coherence block} & \revision{$28$ symbols} \\
        $\mathrm{SINR}_{\subsf UE}$ & Post-beamforming \acs{SINR} per \acs{UE} & \makecell{$[-6,14]$\,dB\\or $[-6,3]$\,dB} \\
        $n_{\subsf sc}$ & Number of active subcarriers & 792 \\
        $\mathrm{SCS}$ & \makecell{Subcarrier spacing\\\revision{(numerology 3)}} & \SI{120}{kHz} \\
        $N_{\subsf DM-RS}$ & \makecell{\revision{Type-2 \acs{DM-RS} \acsp{RE} per resource block} \\ \revision{per code-division multiplexing} \\ \revision{group, 2 symbols}} & \revision{8} \\
        $d_{\subsf UE}$ & \acs{UE} distance from \acs{BS} & \SI{100}{m} to \SI{2}{km} \\
        $v_{\subsf UE}$ & \acs{UE} speed & \makecell{0 to \SI{100}{km/h}\\\revision{(to $120$ when swept)}} \\
        
        \hline
    \end{tabular}
    \label{tab:parameters}
\end{table}

To evaluate the proposed method we conducted ray-tracing simulations with the
NVIDIA Sionna ray tracer~\cite{hoydis2023sionna}---a deterministic,
site-specific channel model rather than a stochastic one---on a map of Denver, USA, with a single \ac{BS} of three sectors, at the
parameters of Table~\ref{tab:parameters} and with the methodology of our
\ac{UMB} studies~\cite{kang2024cellular, jia2025joint}.  The \SI{120}{kHz} \ac{SCS} and \ac{FFT} size of 1024
give a \SI{100}{MHz} channel bandwidth (792 active subcarriers) and make
$T_{\subsf LT} = \SI{10}{ms}$ exactly 80 slots.  \revision{This is a \ac{FR2}
numerology evaluated at an \ac{FR1} carrier: it fixes the \ac{OFDM} grid
arithmetic and the symbol counts of Table~\ref{tab:complexity}, and no
\ac{SINR} reported here depends on it.  Section~\ref{sec:signalling} states the
\ac{SRS} cost in numerology-independent symbols per $T_{\subsf LT}$ and its
\ac{FR1} \SI{30}{kHz} grid fractions.}  In each Monte Carlo \revision{drop} the users of a sector are
placed at random distances, velocities and directions within the ranges of
Table~\ref{tab:parameters}, the \ac{BS} antennas \SI{40}{m} and the \ac{UE}
antennas \SI{1.5}{m} above ground, at a maximum \ac{UE} transmit power of
\SI{26}{dBm} and a \ac{BS} noise figure of \SI{2}{dB}\revision{.  Transmit} powers are then
adjusted so that every connected \ac{UE} falls in the
$[-6,3]$\,dB \ac{SINR} range\revision{.  Long-term} estimation uses
$N_{\subsf SRS} = 8$ \revision{\ac{SRS} snapshots} per \ac{UE} across the assigned resource
blocks.

\begin{figure*}[htbp]
    \centering
    \begin{subfigure}{0.32\textwidth}
        \centering
        \includegraphics[width=\linewidth]{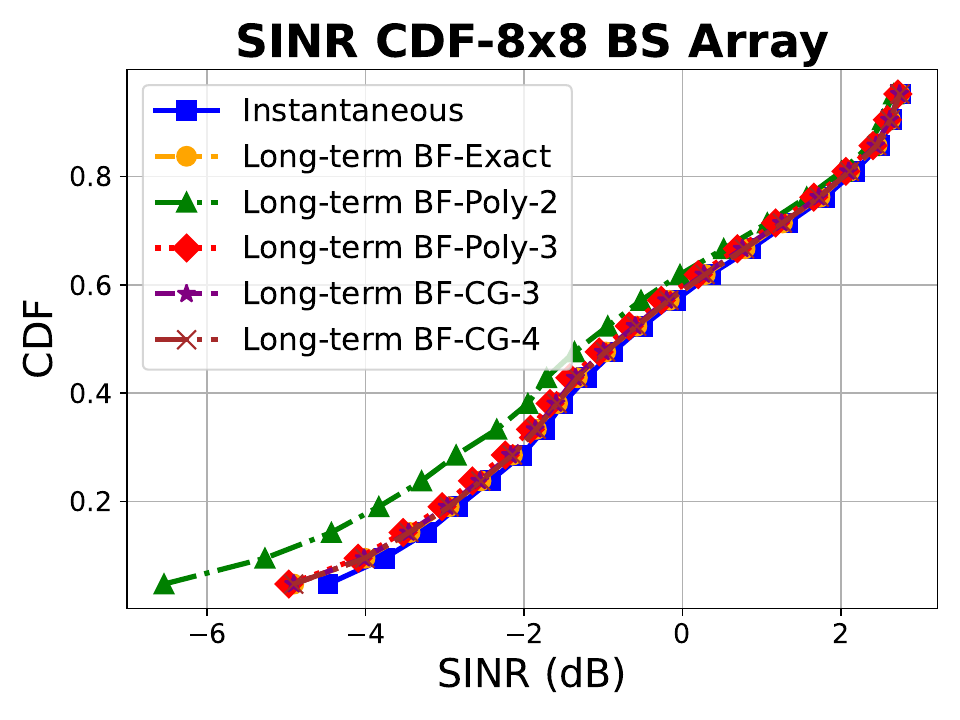}
        \caption{$8\times8$}
        \label{fig:sinr_cdf_8x8_-6-3}
    \end{subfigure}\hfill
    \begin{subfigure}{0.32\textwidth}
        \centering
        \includegraphics[width=\linewidth]{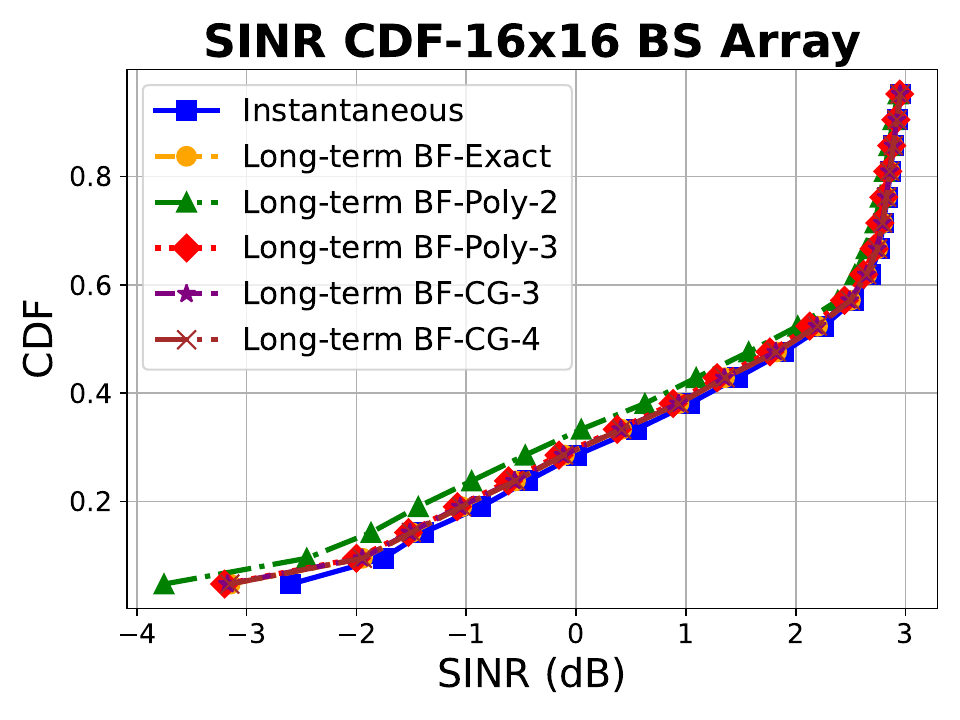}
        \caption{$16\times16$}
        \label{fig:sinr_cdf_16x16_-6-3}
    \end{subfigure}\hfill
    \begin{subfigure}{0.32\textwidth}
        \centering
        \includegraphics[width=\linewidth]{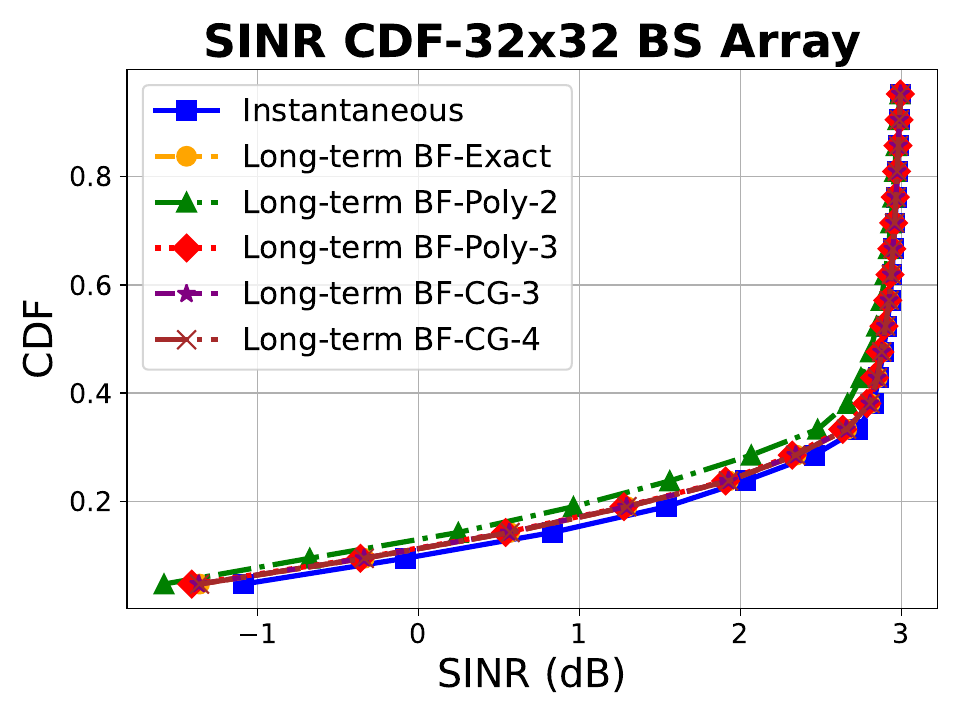}
        \caption{$32\times32$}
        \label{fig:sinr_cdf_32x32_-6-3}
    \end{subfigure}

    \caption{\acp{CDF} of the achievable \ac{SINR}, $10$ \acp{UE} per
    sector, against array size: instantaneous \ac{MMSE}, exact \ac{LTBF},
    order-$d$ polynomial, and $t$-iteration \ac{CG}.}
    \label{fig:sinr_cdf_-6-3}
\end{figure*}

\revision{Every experiment reported in this section
uses single-antenna \acp{UE}, $N_s = 1$, so that
$\wbar{N} = N_{\subsf SRS} = 8$, except the multi-stream paragraph of
Sec.~\ref{sec:rank}, which uses $N_s=2$.  The \ac{SRS} estimates carry additive estimation noise whose
variance sits a factor $G_{\subsf SRS}$ below the data-domain noise floor: a
pilot processing gain, not a received \ac{SNR}, equal to the number of
subcarriers in a coherence bandwidth, since frequency smoothing and the comb's
power concentration trade off exactly; the measured coherence bandwidth of these
channels puts it at $\geq\SI{10.8}{dB}$; we use \SI{10}{dB}.  Section~\ref{sec:coverror} sweeps it upwards from
\SI{0}{dB}, \ac{SRS} at data power with no processing gain;
Fig.~\ref{fig:sinr_cdf_-6-3} and the mechanism sweeps of Secs.~\ref{sec:rank}
and~\ref{sec:ops} use a noiseless \ac{SRS} estimate except where
$G_{\subsf SRS}$ is itself the swept variable, so that each isolates what it
names from covariance-estimation error, and Sec.~\ref{sec:array} reports the
nominal \SI{10}{dB}.  The drops come from three ray-tracing runs of the same
scenes, A--C, and numbers are compared only within a run: Sec.~\ref{sec:coverror}
uses run A; the multi-stream, interferer and \SI{10}{dB}
rank results, Fig.~\ref{fig:sweep_ops}(e) and the
nominal column of Table~\ref{tab:array} use run B; and the rest of
Secs.~\ref{sec:rank}--\ref{sec:array} use run C.  Hence the
$N_s=1$ rural loss at \SI{0}{dB} reads \SI{4.18}{dB} in Sec.~\ref{sec:coverror}
and \SI{4.34}{dB} in Sec.~\ref{sec:rank}.  The pilot-scaling cells of
Sec.~\ref{sec:array} repeat run B's nominal configuration on a second ray
tracing of the scene.  Sec.~\ref{sec:coverror}, Fig.~\ref{fig:sinr_cdf_-6-3},
Table~\ref{tab:array} and the pilot-scaling cells use $100$ drops, and every
other result $500$.}

Two measurements are recorded per drop, one at the start of
$T_{\subsf LT}$, from which the covariances and projections are computed, and
one at its end, at which the \acp{SINR} are measured---the least favorable
comparison against the instantaneous \ac{MMSE} method, since it admits the
greatest channel evolution under user mobility.
\revision{%
Because the two measurements are a full $T_{\subsf LT}$ apart, a \ac{UE} may be in
outage at the first and served at the second, in which case no long-term
projection exists for it.  Such a \ac{UE} is excluded from every reported curve
for that drop---the instantaneous \ac{MMSE} reference included, so that the
two receivers are always scored on an identical set of \acp{UE}---while still
transmitting, so the interference every other \ac{UE} sees is unchanged.  The
rule is invoked for $6$ of the $2377$ connected \ac{UE} instances ($0.25\%$) at
the parameters of Table~\ref{tab:parameters}.  All arithmetic below is double
precision, so no reported gap is due to finite-precision arithmetic; that
case is treated in \cite{kiani2026interference}.}

For each user, three \acp{SINR} are evaluated: the \textbf{instantaneous \ac{MMSE} \ac{SINR}} of~\eqref{eq:Finstant}, an upper bound; the \textbf{exact \ac{LTBF} \ac{SINR}} $\gamma_i^0$, whose projection is \eqref{eq:Giopt} \revision{formed from $\wh{\bs{Q}}_i$ and $\wh{\bs{Q}}$}; and the \textbf{approximate \ac{LTBF} \ac{SINR}} $\gamma_i$ \revision{of \eqref{eq:sinr_proj}}, whose projection uses $\bs{P}(\bs{\beta})$ of~\eqref{eq:Giapprox} or \ac{CG}.

Fig.~\ref{fig:sinr_cdf_-6-3} presents the \acp{CDF} of these quantities over 100 \revision{drops} for three array sizes: the proposed \ac{LTBF} method performs comparably to the instantaneous \ac{MMSE} method \revision{while refreshing its beamforming coefficients three orders of magnitude less often} (Sec.~\ref{sec:complexity}), and a polynomial of order $d=3$ or $t=3$ \ac{CG} iterations already suffices. The accuracy required of $\wh{\bs{Q}}^{-1/2}$ is set by the spread of the \acp{SINR} the base station sees, so widening the target range from $[-6,3]$\,dB to $[-6,14]$\,dB raises the polynomial orders needed for comparable accuracy from 2 and 3 to 10 and 20\revision{.  On the wide target, order $20$ stays within \SI{0.02}{dB} of the exact projection at every array size, and order $10$ loses $0.7$--\SI{1.1}{dB}.  With the noiseless \ac{SRS} estimate of these curves the exact long-term projection lies within \SI{0.1}{dB} of the instantaneous \ac{MMSE} reference at all three array sizes; the losses at the nominal estimator quality are due to covariance estimation, which Sec.~\ref{sec:array} quantifies.}

\revision{%
\subsection{Covariance-Estimation Error and Projection Rank}
\label{sec:coverror}

Assumption (A2) sets $\wh{\bs{Q}}_i = \bs{Q}_i$; the simulator does not.  Every
result in this section is computed from $\wh{\bs{Q}}_i$ of \eqref{eq:QHprod},
formed from $\wbar{N}$ narrowband \ac{SRS} snapshots at a finite processing gain
$G_{\subsf SRS}$, and Fig.~\ref{fig:cov_error} sweeps that estimator over its whole
operating range on the same $100$ drops with common random numbers, with the
exact projection so that no solver error is mixed in.

\emph{Source of the loss.}  With the
(unrealizable) genie covariance the exact long-term projection is a median
\SI{0.28}{dB} below the instantaneous \ac{MMSE} receiver, against \revision{\SI{1.68}{dB} for
the estimator at the nominal $N_{\subsf SRS}=8$ and $G_{\subsf SRS}=\SI{10}{dB}$
and \SI{4.18}{dB} at \SI{0}{dB}}: every larger gap below is estimation error.

\emph{Processing gain and snapshot count.}  At $G_{\subsf SRS} = \SI{0}{dB}$
raising $N_{\subsf SRS}$ from $2$ to $32$, $16$ times the snapshots from the
same sounding, gains \SI{3.6}{dB}, as much as raising $G_{\subsf SRS}$ from $0$ to
$\SI{20}{dB}$ at $N_{\subsf SRS}=8$, whereas at $\SI{20}{dB}$ the whole
$N_{\subsf SRS}$ sweep gains only \SI{0.68}{dB} and the loss against the
instantaneous receiver is \SI{0.55}{dB}.  The per-snapshot processing gain
therefore matters more than the number of snapshots, and the \SI{0}{dB} end of
the sweep is a bound, not an operating point.

\emph{Rank inflation.}  Panel~(b)
plots $\tr(\wh{\bs{Q}}_i)^2/\tr(\wh{\bs{Q}}_i^2)$, which reaches $1.8$, $3.1$,
$4.8$, $6.7$ and $8.2$ at $N_{\subsf SRS} = 2,4,8,16,32$ and $\SI{0}{dB}$ where
the true $\bs{Q}_i$ is effectively rank one, falling to $1.0$--$1.1$ at $\SI{20}{dB}$.
The trailing eigenvalues thus come from estimation noise rather than
propagation, and a rank-$r$ projection built from such an estimate devotes some
of its $r$ dimensions to them.  Ledoit--Wolf shrinkage of \eqref{eq:QHprod}
gains \SI{1.64}{dB} at $N_{\subsf SRS}=2$, is neutral at $8$ and loses
$0.3$--\SI{0.4}{dB} at $16$ and $32$, where the sample covariance is already well
conditioned; it is therefore useful only with few snapshots.

\emph{Projection rank.}  Panel~(c) sweeps $r$ with a noiseless estimate
and the median \ac{SINR} rises only from \SI{4.82}{dB} at $r=1$ to \SI{4.86}{dB} at
$r=2$ and \SI{4.90}{dB} at $r=8$: with a single dominant path per \ac{UE}
(Sec.~\ref{sec:assumptions}) there is no second propagation direction for $r>1$
to capture.  Hence $r=1$ loses only \SI{0.04}{dB} relative to the $r=2$ used elsewhere
while removing the per-resource-element penalty against \eqref{eq:Fgram}
entirely (Sec.~\ref{sec:complexity}), and the design rule for $r$ must come
from a scenario with appreciable angular spread, which Sec.~\ref{sec:rank}
supplies.

\begin{figure*}[t]
    \centering
    \includegraphics[width=\linewidth]{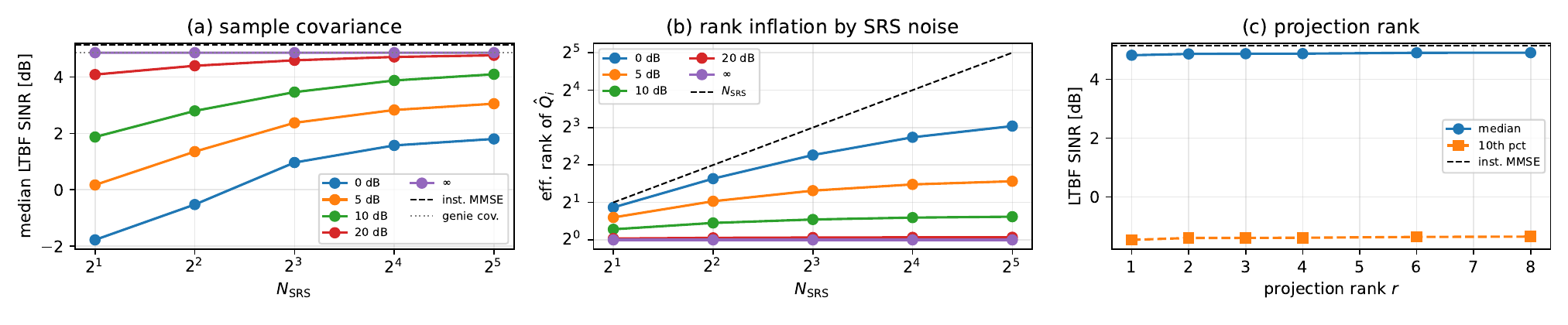}\\[2pt]
    \includegraphics[width=0.6667\linewidth]{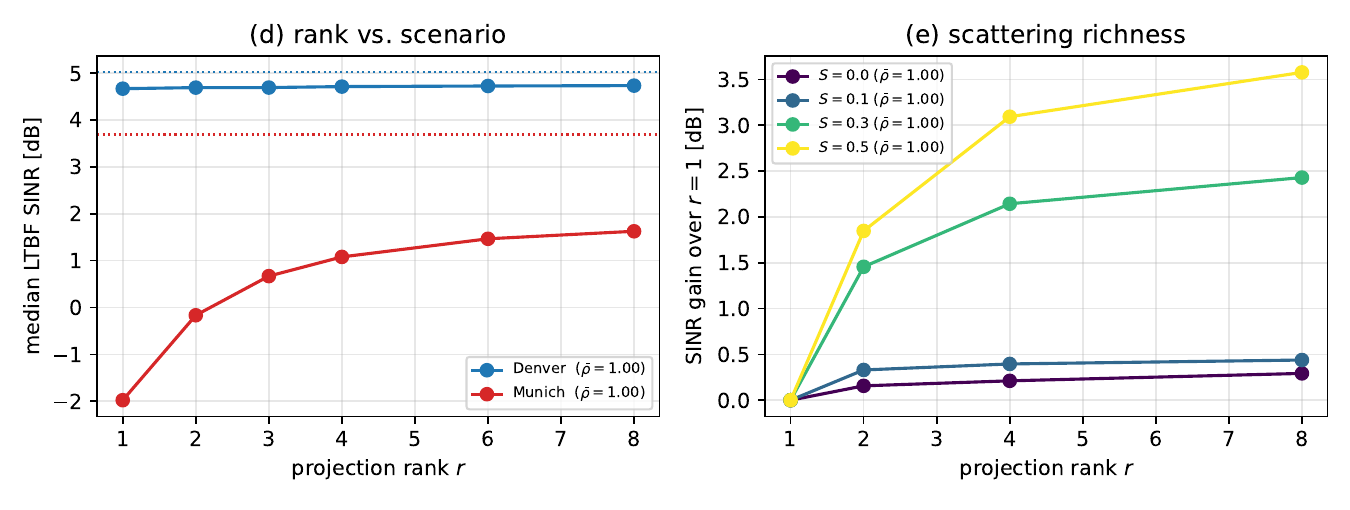}
    \caption{\revision{Covariance-estimation and projection-rank sweeps for an
    $8\times8$ array using the exact \ac{LTBF} projection. Rural Denver,
    $100$ drops: (a)~median \ac{LTBF} \ac{SINR} and (b)~effective rank of
    $\wh{\bs{Q}}_i$ versus the number of \ac{SRS} snapshots, with curves for
    processing gain $G_{\subsf SRS}$; (c)~median and 10th-percentile \ac{LTBF}
    \ac{SINR} versus projection rank $r$ with a noiseless estimate. With
    $500$ drops and noiseless estimates: (d)~median \ac{LTBF} \ac{SINR} versus
    $r$ in rural Denver and dense-urban Munich; (e)~\ac{SINR} gain over $r=1$
    versus $r$ in Munich for material scattering coefficients $S$.}}
    \label{fig:cov_error}%
    \label{fig:sweep_rank}
\end{figure*}
}

\revision{%
\subsection{Multipath Scenario and Rank Selection}
\label{sec:rank}

Section~\ref{sec:coverror} showed that no design rule for $r$ can be drawn
from a scenario in which $\bs{Q}_i$ is effectively rank one.  This subsection
supplies one, from a second scenario built for the purpose.  Every result in
this subsection and the next uses $500$ drops, is computed with a noiseless \ac{SRS}
estimate unless another $G_{\subsf SRS}$ is named, and reports the
median over the \acp{UE} the scheduling rule serves, so compared curves are
always averaged over the same population.

\emph{Diagnosis.}  Table~\ref{tab:scenarios} reports, per \ac{UE}, the
second-to-first singular-value ratio of $\bs{H}_i$ and the effective rank
$\bar{\rho} = \tr(\bs{Q}_i)^2/\tr(\bs{Q}_i^2)$.  In the rural Denver scenario the median
ratio is $6.7\times10^{-6}$ and \revision{only $0.26\%$ of} \acp{UE} have $\bar{\rho}>1.1$: the
scene resolves a single dominant path per \ac{UE}, which is why $r=2$ suffices
there.
A denser map alone does not make the rank matter.  We use the Munich scene of~\cite{hoydis2023sionna}.  Three changes
were each necessary and none alone sufficient: the \ac{BS} must sit at
\SI{25}{m}, below the rooftops, or the \ac{LoS} ray dominates and $\bar{\rho}$
stays at one; diffuse reflection, disabled by default in the ray tracer, must be
enabled; and the scattering coefficient $S$, zero for every material of the
default scene, must be raised into the range $0.2$--$0.5$ that
\mbox{ITU-R~P.2040} assigns to building surfaces.  Table~\ref{tab:scenarios}
sweeps $S$ in one fixed geometry; at $S=0.5$ the scene is the dense-urban
scenario of the rest of this section.  The $S=0$ row of Table~\ref{tab:scenarios} is the control---the
single-path propagation model evaluated in the harsher geometry---and on the
$10\%$ of \acp{UE} it links its rank behavior is close to the rural scenario's, $r=1\to8$ gaining
\SI{0.29}{dB} against \SI{0.07}{dB}, so the second eigenvalue is created by the
scattering mechanism.

\emph{Rank sweep and design rule.}
Fig.~\ref{fig:sweep_rank}(d) sweeps $r$ in both scenarios: the median \ac{LTBF}
\ac{SINR} spans \SI{0.07}{dB} in the rural scenario (\SI{4.67}{} to \SI{4.73}{dB}) and
\SI{3.60}{dB} in the dense-urban one (\SI{-1.98}{} to \SI{1.63}{dB}), and the loss against
the instantaneous \ac{MMSE} receiver at $r=2$ grows from \SI{0.32}{dB} to
\SI{3.86}{dB}---$r=2$ is close to sufficient when the channel has one
dominant direction and visibly insufficient when it does not.
Fig.~\ref{fig:sweep_rank}(e) plots the gain over $r=1$ at each scattering
strength, and it tracks the fraction of \acp{UE} with $\bar{\rho}>1.1$
monotonically---$0.29$, $0.44$, $2.43$ and \SI{3.58}{dB} at $0.21\%$, $1.4\%$,
$14\%$ and $22\%$ of such \acp{UE}---so the
rank should be chosen from the measured effective rank of $\wh{\bs{Q}}_i$, a
quantity the base station already forms.  The returns diminish steadily: in the dense-urban scenario the marginal gain
per unit of $r$ falls from \SI{1.81}{dB} between $r=1$ and $2$ to \SI{0.08}{dB}
between $r=6$ and $8$, against an instantaneous complexity growing as
$\mathcal{O}(r^3)$.

\emph{Robustness.}  The contrast survives estimation error: at $G_{\subsf SRS}=\SI{10}{dB}$ the rural span
widens only to \SI{0.24}{dB} while the dense-urban one is still
\SI{2.07}{dB}.  What little the rank does in the rural scenario is due to
the estimation noise that inflates the effective rank of $\wh{\bs{Q}}_i$
(Sec.~\ref{sec:coverror}); with that noise removed the rank has almost no
effect where the channel has one direction, as expected for a rank-one $\bs{Q}_i$.

\emph{Multi-stream reception.}  We take $N_s=2$ on a $1\times2$ \ac{UE}
array, with $r=2$, the smallest rank the constraint $r\geq N_s$ permits, and
$G_{\subsf SRS}=\SI{0}{dB}$, and measure the per-stream \ac{MMSE} \ac{SINR}
$1/[(\bs{I}+\alpha_i\wt{\bs{H}}_i\herm\wt{\bs{R}}_i^{-1}\wt{\bs{H}}_i)^{-1}]_{ss}-1$
of stream $s$.  Relative to the instantaneous \ac{MMSE} receiver, the loss of
the long-term projection is half that at $N_s=1$ in the rural scenario
(\SI{2.22}{dB} against \SI{4.34}{dB}) and $40\%$ lower in the dense-urban one
(\SI{4.36}{dB} against \SI{7.24}{dB}), and $r=8$ recovers a further $0.5$ and
\SI{1.7}{dB}.  The reduced observation space carries more signal and less
interference, so the truncation costs proportionally less, and
the long-term architecture is more favorable in the multi-stream setting than in
the single-stream one.

\begin{table}[t]
\centering
\caption{Propagation statistics, per \ac{UE} over the strongest sector and all $69$ resource blocks; \emph{linked} is the fraction of dropped \acp{UE} the ray tracer connects at all. $S=0.1$ and $S=0.3$ are swept in Fig.~\ref{fig:sweep_rank}(e).}
\label{tab:scenarios}
\scriptsize
\setlength{\tabcolsep}{2.5pt}
\begin{tabular}{@{}lcccccc@{}}
\toprule
Scenario & drops & linked & $\sigma_2/\sigma_1$ & $\bar{\rho}_{90}$ & $\bar{\rho}_{\max}$ & $\bar{\rho}>1.1$ \\
\midrule
Rural (Denver) & 500 & 79\% & $6.7\times10^{-6}$ & 1.00 & 1.85 & 0.26\% \\
Munich, $S=0$ (control) & 500 & 10\% & $<\!10^{-12}$ & 1.00 & 1.52 & 0.21\% \\
Dense-urban (Munich, $S=0.5$) & 500 & 81\% & $0.036$ & 1.45 & 2.76 & 22\% \\
\bottomrule
\end{tabular}
\end{table}

}

\revision{%
\subsection{Operating Conditions: Mobility, Load and Clustering}
\label{sec:ops}

Fig.~\ref{fig:sweep_ops} sweeps the conditions under which the projection must
remain accurate.  Mobility and the update period enter only through the displacement
$v\,T_{\subsf LT}$ between the measurement from which the statistics are computed
and the one at which the \ac{SINR} is measured, so panels~(a) and~(b) fix one
and sweep the other.

\emph{Update period and mobility.}  Sweeping $T_{\subsf LT}$ from \SI{1}{ms} to
\SI{100}{ms} at the nominal speed distribution moves the median \ac{SINR} by
about \SI{0.5}{dB} with no monotone trend (the extremes are
\SI{5.01}{dB} at \SI{5}{ms} and \SI{4.48}{dB} at \SI{100}{ms}), so the
$1.3\times10^{3}$ reduction in refresh rate at the $r=2$ used in this sweep does
not degrade tracking, and a deployment may lengthen $T_{\subsf LT}$ well beyond
\SI{10}{ms}.  At fixed $T_{\subsf LT}=\SI{10}{ms}$, raising the \ac{UE} speed
from $0$ to \SI{120}{km/h} costs \SI{0.10}{dB} in the rural scenario, within the
run-to-run spread, and \SI{2.74}{dB} in the dense-urban one, monotonically.  A
single-path signature $\bs{u}_i$ rotates only as fast as the angle to
the \ac{BS} changes, whereas several comparable paths decorrelate as their
relative phases evolve.  The
design rule is that $T_{\subsf LT}$ should be tied to the angular coherence
time rather than the small-scale one, and shortened as the scattering becomes
richer.

\emph{System load.}
As $N_{\subsf UE}$ is swept from $5$ to $40$ per sector
(Fig.~\ref{fig:sweep_ops}(c)), i.e., up to $N_{\subsf UE}r = 80$ observed
dimensions against $N_{\subsf rx}=64$ antennas, the loss against instantaneous
\ac{MMSE} falls from \SI{0.23}{} to \SI{0.09}{dB} in the rural scenario,
negligible at either end, and from \SI{3.90}{} to \SI{2.64}{dB} in the
dense-urban one; the long-term projection is thus most competitive in heavily
loaded sectors, where the instantaneous receiver is most expensive.

\emph{Effect of whitening.}  The projection of \eqref{eq:Giapprox} differs from
classical statistical-\ac{CSI} beamforming only in whitening by the sector
covariance $\bs{Q}^{-1/2}$ before truncating, so panels~(d) and~(e) compare it
with \emph{eigen-beamforming}, whose $\bs{G}_i$ is the
top-$r$ eigenvectors of $\wh{\bs{Q}}_i$, and \emph{\ac{DFT} beamspace
selection}, whose $\bs{G}_i$ is the $r$ strongest beams of the array's $2$-D
\ac{DFT} dictionary.  These two reference projections use the same $\wh{\bs{Q}}_i$ from the same
\ac{SRS} realization and are scored by the same reduced-domain receiver, and
the proposed projection reduces to eigen-beamforming when $\bs{Q}$ is a
multiple of the identity, so the gap between those two is attributable to
interference statistics alone.  Against
eigen-beamforming the proposed projection is ahead by $1.35$, $1.51$, $1.57$,
$1.67$ and \SI{1.78}{dB} at $N_{\subsf UE}=5$, $10$, $20$, $30$ and $40$, the
margin widening with the load as there is more interference for
$\bs{Q}^{-1/2}$ to suppress.  At $N_{\subsf UE}=10$,
it trails eigen-beamforming by \SI{0.32}{dB} at the
\SI{0}{dB} \ac{SRS} lower bound but leads by $0.38$, $0.77$ and \SI{1.32}{dB}
at the nominal \SI{10}{dB}, at \SI{20}{dB} and with a noiseless estimate
\revision{(on run B, hence not the \SI{1.51}{dB} of the load sweep)}.  The
reference projections are nearly flat across that sweep (\SI{1.61}{} and
\SI{0.38}{dB}) because neither uses the covariance for anything but a subspace,
whereas the proposed projection gains \SI{3.25}{dB}.

\emph{Design rule for whitening.}  Whitening by $\bs{Q}^{-1/2}$ helps when the
covariance estimate is accurate enough, and helps more as interference rather
than noise limits the sector; the threshold measured here is
$G_{\subsf SRS}\approx\SI{5}{dB}$, below the nominal operating point of
Table~\ref{tab:parameters}.  Below it a
statistical-\ac{CSI} projection without whitening is the better choice.

\emph{Clustering and interference.}  Both experiments use the dense-urban scenario, each
measured against its own uniform-drop reference.  Drawing the \acp{UE} of a sector from two discs of
\SI{60}{m} radius instead of uniformly makes their long-term subspaces overlap,
the hardest case for any statistical-\ac{CSI} projection; it costs a
near-constant \SI{1.42}{}--\SI{1.50}{dB} across $r=1,2,4,8$
(Fig.~\ref{fig:sweep_ops}(f)); the penalty is independent of $r$ because
clustering removes signal separation, which no projection rank can restore.  A \SI{30}{dB} out-of-sector interferer costs \revision{(at $G_{\subsf SRS}=\SI{0}{dB}$)} \SI{1.76}{dB} at $r=1$
against \SI{0.94}{dB} at $r=8$, the opposite trend, because nulling an
interferer needs a spare observation dimension; the conditioning it induces is
studied in \cite{kiani2026interference}.

\begin{figure*}[t]
    \centering
    \includegraphics[width=\linewidth]{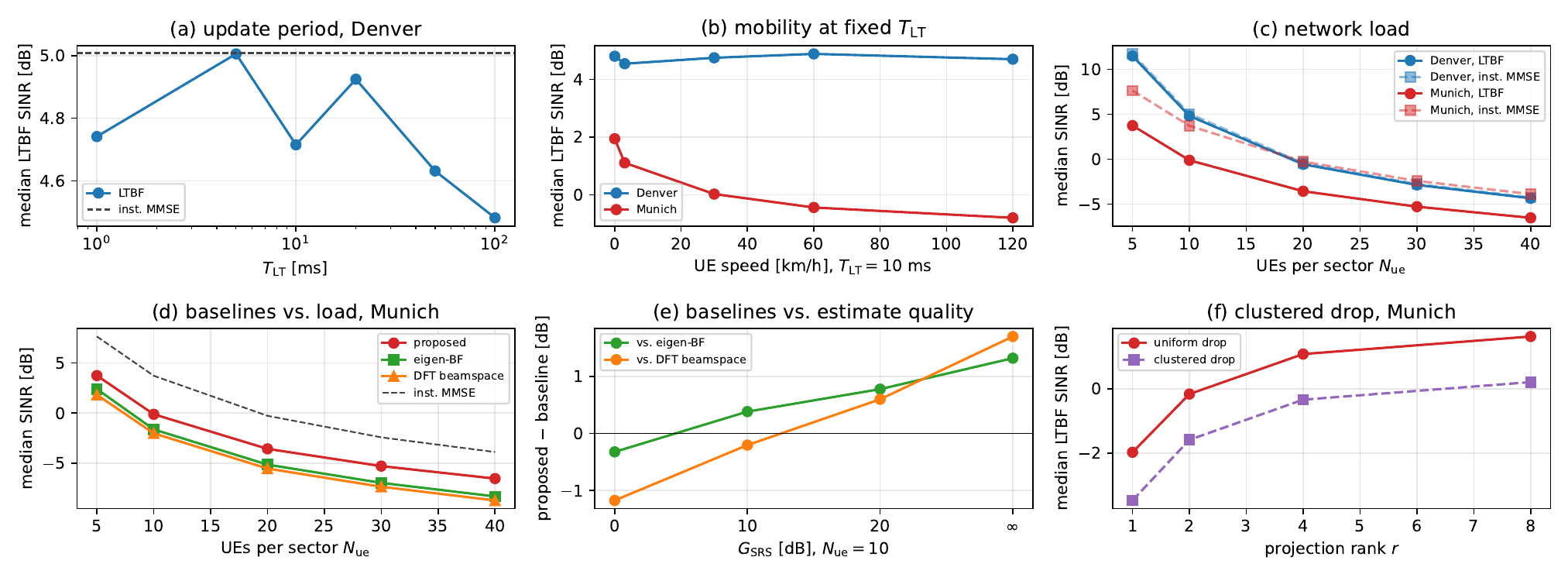}
    \caption{\revision{Operating conditions, $r=2$ except in~(f):
    (a)~$T_{\subsf LT}$, (b)~\ac{UE} speed, (c)~$N_{\subsf UE}$, (d)~load
    and (e)~estimate quality against the two reference projections, (f)~clustered drop.
    (a) is rural, (d)--(f) dense-urban, (b) and (c) both.}}
    \label{fig:sweep_ops}
\end{figure*}

}

\revision{%
\subsection{Array Size and Pilot-Budget Scaling}
\label{sec:array}

Array size interacts with the covariance estimator, and
Table~\ref{tab:array} separates the two by evaluating every array size with
the estimator of \eqref{eq:QHprod} at the nominal $G_{\subsf SRS}$ and with a
noiseless \ac{SRS} estimate, the nominal column on run B.

\emph{Loss against array size.}  At the
nominal $G_{\subsf SRS}=\SI{10}{dB}$ and $N_{\subsf SRS}=8$ the loss against
the instantaneous receiver is $2.25$, $6.30$ and \SI{10.58}{dB} for the
$[-6,3]$ target as the array grows.  The estimate $\wh{\bs{Q}}_i$ is built from
$\wbar{N} = N_sN_{\subsf SRS}$ columns however large $N_{\subsf rx}$ is, but
the dominant mechanism is that uplink power control caps the total
received \ac{SNR} at the upper end of the target range.  A \ac{UE} that would
exceed it is backed off, so its total received energy is held fixed while
$N_{\subsf rx}$ grows and its per-antenna pilot energy falls as
$1/N_{\subsf rx}$.  Because $G_{\subsf SRS}$ is a gain against an absolute
noise floor, the per-antenna estimation quality of these \acp{UE} degrades with
the array.  The measurement matches: each fourfold increase in
$N_{\subsf rx}$ is a \SI{6}{dB} drop in per-antenna pilot energy and costs
$4.05$ and \SI{4.28}{dB}.  The prediction that distinguishes this mechanism
from a generic large-array effect is that the wider $[-6,14]$ target, where far
fewer \acp{UE} are backed off, should degrade much less, and it does: it
loses $1.76$, $3.07$ and \SI{2.95}{dB} and saturates.

\emph{Pilot scaling.}
Raising $N_{\subsf SRS}$ from $8$ to $32$ at the nominal $G_{\subsf SRS}$
gains \SI{1.40}{dB} at $8\times8$ and \SI{3.87}{dB} at $16\times16$ for the
$[-6,3]$ target, reducing the loss against the instantaneous receiver
within that sweep to $0.86$ and \SI{2.49}{dB} respectively (the
sweep has its own ray tracing, so its gains are not compared with
Table~\ref{tab:array}).  That the recovery is $2.8\times$ larger for the larger array is the
signature of the mechanism above.
The design rule is therefore explicit: $N_{\subsf SRS}$ should grow with
$N_{\subsf rx}$, which costs estimation arithmetic rather than \ac{SRS} resources.  Two further headroom
terms---multi-symbol \ac{SRS}, and spreading the $N_{\subsf SRS}$ snapshots
across coherence bandwidths wide enough to make them statistically independent---are available and are not claimed here.

\begin{table}[t]
\centering
\caption{Median \ac{LTBF} \ac{SINR} in dB against array size; $100$ drops, $r=2$, exact projection. \emph{Nominal} and \emph{noiseless} are \eqref{eq:QHprod} at $G_{\subsf SRS} = \SI{10}{dB}$ and with no \ac{SRS} estimation noise. \revision{\emph{Noiseless} and \emph{inst.\ \acs{MMSE}} use run C and \emph{nominal} run B (Sec.~\ref{sec:results}); \emph{loss} is nominal against the instantaneous \ac{MMSE} of its own run.}}
\label{tab:array}
\footnotesize
\begin{tabular}{lccccc}
\toprule
$N_{\subsf rx}$ & target & nominal & noiseless & inst.\ \acs{MMSE} & loss \\
\midrule
$8\times8$ & $[-6,3]$ & $-2.89$ & $-0.79$ & $-0.70$ & $2.25$ \\
$8\times8$ & $[-6,14]$ & $3.21$ & $4.88$ & $5.10$ & $1.76$ \\
$16\times16$ & $[-6,3]$ & $-4.11$ & $2.00$ & $2.06$ & $6.30$ \\
$16\times16$ & $[-6,14]$ & $9.06$ & $11.63$ & $11.93$ & $3.07$ \\
$32\times32$ & $[-6,3]$ & $-7.67$ & $2.89$ & $2.90$ & $10.58$ \\
$32\times32$ & $[-6,14]$ & $10.84$ & $13.75$ & $13.78$ & $2.95$ \\
\bottomrule
\end{tabular}
\end{table}

}

\subsection{Hyperparameter Selection and Sensitivity}
\label{sec:hyper}

The three hyperparameters trade complexity against performance.  The
rank $r$ follows the measured effective rank of $\wh{\bs{Q}}_i$
(Sec.~\ref{sec:rank}) at an instantaneous per-\ac{UE} cost
$\mathcal{O}(r^3)$; the order $d$ and \ac{CG} count $t$ follow the accuracy
needed in $\wh{\bs{Q}}^{-1/2}$, hence the eigenvalue spread of $\wh{\bs{Q}}$
and, through Proposition~\ref{prop:error}, the dynamic range of the received
\acp{SINR}, \revision{the count scaling with
$\sqrt{\mathrm{cond}(\wh{\bs{Q}})}$ and its remedies being pursued in
\cite{kiani2026lowrank}}.  The period $T_{\subsf LT}$ must be shorter than the
large-scale coherence time (Sec.~\ref{sec:ops}) yet long enough to amortize the
centralized computation.

\section{Conclusion}
\label{sec:conclusion}

\revision{%
The projection is available in closed form from an inverse matrix square root;
it maximizes a capacity upper bound and exactly maximizes the mean
post-projection \ac{SINR}, though not the ergodic capacity itself.  The savings this
architecture delivers depend on which realization of the instantaneous
\ac{MMSE} receiver is taken as the baseline.  Against the
full-dimensional form \eqref{eq:Finstant} the proposed method reduces the
operation count by more than two orders of magnitude.  Against the
algebraically identical Gram-domain form \eqref{eq:Fgram}, in which the \ac{MRC}
reduction precedes an $N_{\subsf UE}N_s$-dimensional inverse, the proposed
method needs more operations at the $10$-user configuration
of Table~\ref{tab:complexity} and becomes cheaper only above $N_{\subsf UE}N_s \approx 15$
single-stream \acp{UE} with the \ac{CG} solve at $r=1$.  It does, however,
reduce the rate at which beamforming coefficients must be acquired and
refreshed, by three orders of magnitude at that operating point, and it
removes the coupling between users in the real-time path; these are the binding
constraints for extreme arrays.

Three limitations remain.  First, the per-resource-element cost of the
projection exceeds that of \ac{MRC} by the factor $r/N_s$ whenever $r>N_s$.  Second, the
rural scenario evaluated here resolves a single dominant path per
user, so $\bs{Q}_i$ is effectively rank one for all but $0.26\%$ of users
(Sec.~\ref{sec:rank}).  In this scenario, $r>1$ helps only through the mismatch between the
model and the channel, namely the noise-inflated rank of the estimated
covariance and the channel's evolution over $T_{\subsf LT}$
(Sec.~\ref{sec:assumptions}), so the design rule for $r$, choosing it from the
measured effective rank of $\wh{\bs{Q}}_i$, comes from the dense-urban scenario of
Sec.~\ref{sec:rank}.  Third, the results assume the long-term statistics are
stationary over $T_{\subsf LT}$, an assumption that weakens for extremely large
arrays exhibiting spatial non-stationarity.  The hardware continuation, namely
the finite-precision behavior of these inversions under strong interference
\cite{kiani2026interference} and their acceleration by low-rank preconditioning
in the beamspace domain \cite{kiani2026lowrank}, is reported separately.}

\appendices
\section{Proof of Lemma~\ref{lem:capupper}} \label{sec:proof_projection}

Assume $\bs Q \succ 0$ and for any potential 
projection matrix $\bs{G}_i$, 
define the whitened matrix $\bs{G}_i\wtn := \bs G_i \, \bs Q^{1/2}$.
Take a thin QR decomposition of $\bs{G}_i\wtnh$:
\begin{equation} \label{eq:BVT}
    \bs{G}_i\wtnh \;=\; \bs V_i \, \bs T_i,
\end{equation}
where $\bs V_i \in \mathbb{C}^{N_{\subsf rx}\times r}$ has orthonormal columns
($\bs V_i\herm\bs V_i=\bs I_r$), and
$\bs T_i \in \mathbb{C}^{r\times r}$ is an invertible upper triangular matrix.
Let
\begin{equation} \label{eq:Qitilde_def}
    \bs{Q}_i\wtn \;:=\; \bs Q^{-\subsf H/2}\, \bs Q_i \, \bs Q^{-1/2} \succeq \bs 0,
    \qquad
    \bs{\Upsilon}_i \;:=\; \bs V_i\herm \bs{Q}_i\wtn \bs V_i \succeq \bs 0.
\end{equation}
The two factors in $\Lambda_i(\bs G_i)$ 
in \eqref{eq:capupper} are:
\begin{align}
    &\bs G_i \bs Q_i \bs G_i\herm
      \;=\; \bs T_i\herm \, \bs{\Upsilon}_i \, \bs T_i,
      \label{eq:num_block}
      \\
    &\tilde{\bs{R}}_i = \bs G_i (\bs Q - \alpha_i \bs Q_i)\bs  G_i\herm
      \;=\; \bs T_i\herm \, (\bs I_r - \alpha_i \bs{\Upsilon}_i)\, \bs T_i,
      \label{eq:den_block}
\end{align}
where we used
$\bs Q - \alpha_i \bs Q_i
 = \bs Q^{\subsf H/2}(\bs I - \alpha_i \bs{Q}_i\wtn)\bs Q^{1/2}$,
$\bs V_i\herm\bs V_i=\bs I_r$, and
\revision{$\bs{G}_i\wtn\bs{G}_i\wtnh = \bs G_i \bs Q^{\subsf H/2}\bs Q^{1/2}\bs
G_i\herm = \bs G_i\bs Q\bs G_i\herm$.  Both identities are congruences by a
Hermitian factor, $\bs{Q}_i\wtn$ being $\bs Q_i$ conjugated by $\bs Q^{-1/2}$.}
Therefore,
\begin{equation}
\begin{aligned}
    \Lambda_i(\bs G_i)
    \;&=\; \big(\bs T_i\herm\bs{\Upsilon}_i \bs T_i\big)\,
           \Big[\bs{T}_i\herm(\bs{I}_r - \alpha_i \bs{\Upsilon}_i)\bs T_i\Big]^{-1}
\\[2pt]
    \;&=\; \bs{T}_i\herm\,
            \bs{\Upsilon}_i(\bs I_r - \alpha_i \bs{\Upsilon}_i)^{-1}\,
            (\bs T_i\herm)^{-1},
\end{aligned}
\end{equation}
and the capacity upper bound in 
\eqref{eq:capupper} is
\begin{align}
    \MoveEqLeft \wbar{C}_i(\bs{G}_i) 
    = 
    \log_2 \det\!\big(\bs I_r + \revision{\alpha_i}\Lambda_i(\bs G_i)\big) \nonumber \\
    &=
    \log_2 \det\!\Big(\bs{I}_r + \revision{\alpha_i}\bs{\Upsilon}_i(\bs{I}_r - \alpha_i \bs{\Upsilon}_i)^{-1}\Big),
    \label{eq:C_logdet}
\end{align}
where we have used that $\det( \bs{I} + \bs{T} \bs{Z} \bs{T}^{-1}) = \det (\bs{I} +  \bs{Z})$
for any invertible $\bs{T}$.
\revision{Equation \eqref{eq:C_logdet} depends on $\bs{G}_i$ only through
$\bs{\Upsilon}_i$, hence only through the subspace spanned by $\bs{V}_i$ and not
through $\bs{T}_i$, which is the invariance stated in Lemma~\ref{lem:capupper}.}
Let $\{\mu_k\}_{k=1}^r$ be the eigenvalues of $\bs{\Upsilon}_i=\bs V_i\herm\bs{Q}_i\wtn \bs V_i$
(the Ritz values of $\bs{Q}_i\wtn$ on the $r$-dimensional subspace spanned by $\bs V_i$).
From \eqref{eq:C_logdet},
\begin{equation}
\label{eq:C_sum}
    \wbar{C}_i(\bs G_i)
    \;=\;
    \sum_{k=1}^r
    \log_2\!\Big(1 + \frac{\revision{\alpha_i}\mu_k}{1-\alpha_i \mu_k}\Big),
    \qquad
    0 \le \mu_k < \tfrac{1}{\alpha_i}.
\end{equation}
Since $\bs{Q} - \alpha_i \bs{Q}_i = I + \sum_{j \neq i} \alpha_j \bs{Q}_j \succ 0$, \revision{congruence by $\bs{Q}^{-1/2}$} yields $I - \alpha_i \bs{Q}_i\wtn \succ 0$, so all eigenvalues of $\bs{Q}_i\wtn$ are strictly less than $1/\alpha_i$; by Poincaré's separation theorem\revision{~\cite[Cor.~4.3.37]{horn2012matrix}} the Ritz values inherit $0 \le \mu_k < 1/\alpha_i$.
The map $\mu \mapsto \revision{\alpha_i}\mu/(1-\alpha_i\mu)$ is strictly increasing on
$[0,1/\alpha_i)$, so the sum in \eqref{eq:C_sum} is strictly increasing in each
$\mu_k$, and by the Ky Fan/Courant--Fischer
principle\revision{~\cite[Ch.~4]{horn2012matrix}} it is maximized when $\bs V_i$ spans the $r$
principal eigenvectors of $\bs{Q}_i\wtn$, i.e., when $\bs{V}_i\herm = [
\bs{Q}_i^{1/2}\bs{Q}^{-1/2}]_r$, the eigenvectors of $\bs{Q}_i\wtn$ being the
right singular vectors of $\bs{Q}_i^{1/2}\bs{Q}^{-1/2}$.
Since $\wbar{C}_i(\bs{G}_i)$ is independent of $\bs{T}_i$ we may take $\bs{T}_i
= \bs{I}_r$ in \eqref{eq:BVT}, so that $\bs{G}_i\wtn = \bs{V}_i\herm$ and the
corresponding maximizing $\bs{G}_i = \bs{G}_i\wtn\bs{Q}^{-1/2}$ is
\eqref{eq:Giopt}.

Finally, we prove Remark~\ref{rem:Gioptfact}. Suppose $\bs Q_i$ has a factorization $\bs{Q}_i = c\, \bs{B}_i \bs{B}_i\herm$ for a matrix $\bs{B}_i$ and scalar $c > 0$, so that $\bs{Q}_i\wtn = c\, (\bs{B}_i\herm \bs{Q}^{-1/2})\herm (\bs{B}_i\herm \bs{Q}^{-1/2})$.
The non-zero eigenvalues and corresponding eigenvectors of a positive semidefinite matrix $\bs{Z}\herm \bs{Z}$ are identical to the squared non-zero singular values and corresponding right singular vectors of $\bs{Z}$. Letting $\bs{Z} = \bs{B}_i\herm \bs{Q}^{-1/2}$, the principal eigenvectors of $\bs{Q}_i\wtn$ are the top $r$ right singular vectors of $\bs{B}_i\herm \bs{Q}^{-1/2}$ (the scalar $c > 0$ does not change the eigenvectors). Following the same reasoning as above, the optimal subspace is spanned by $\bs{V}_i\herm = [\bs{B}_i\herm \bs{Q}^{-1/2}]_r$, and substituting yields $\bs{G}_i = \bs{V}_i\herm \bs{Q}^{-1/2} = [\bs{B}_i\herm \bs{Q}^{-1/2}]_r \bs{Q}^{-1/2}$, completing the proof.

\revision{%
\emph{Proof of Corollary~\ref{cor:sinr}.}\ \
Let $N_s = 1$, so $\tilde{\bs{h}}_i = \bs{G}_i\bs{h}_i$ is an $r$-vector and
$\tilde{\bs{R}}_i = \bs{G}_i\bs{R}_i\bs{G}_i\herm$ is deterministic, being a
function of the long-term statistics only.  Taking
the expectation of \eqref{eq:sinr_proj} over the small-scale fading and using
the cyclic property of the trace,
\begin{align}
    \Exp\!\left[\gamma_i(\bs{G}_i)\right]
    &= \alpha_i\,\Exp\!\left[\tr\!\big(\tilde{\bs{R}}_i^{-1}
       \tilde{\bs{h}}_i\tilde{\bs{h}}_i\herm\big)\right] \nonumber \\
    &= \alpha_i\,\tr\!\big(\tilde{\bs{R}}_i^{-1}\bs{G}_i
       \Exp[\bs{h}_i\bs{h}_i\herm]\bs{G}_i\herm\big) \nonumber \\
    &= \alpha_i\,\tr\!\big(\bs{G}_i\bs{Q}_i\bs{G}_i\herm
       \tilde{\bs{R}}_i^{-1}\big)
     = \alpha_i \tr\!\big(\Lambda_i(\bs{G}_i)\big),
\end{align}
by the definitions \eqref{eq:Qjdef} and \eqref{eq:capupper}.  No inequality has been
used: unlike $\wbar{C}_i$, the mean \ac{SINR} is computed exactly.  Substituting
the whitened form of $\Lambda_i$ derived above and invoking the similarity
invariance of the trace gives $\tr(\Lambda_i) =
\tr\big(\bs{\Upsilon}_i(\bs I_r - \alpha_i\bs{\Upsilon}_i)^{-1}\big)
= \sum_{k=1}^r \mu_k/(1-\alpha_i\mu_k)$, which is \eqref{eq:Egamma}.  This is
strictly increasing in each Ritz value $\mu_k$ on $[0,1/\alpha_i)$, exactly as
\eqref{eq:C_sum} is, so the maximization over $r$-dimensional subspaces is the
identical problem and the Ky Fan/Courant--Fischer principle again selects the
span of the $r$ principal eigenvectors of $\bs{Q}_i\wtn$, giving
\eqref{eq:Giopt}.  \qed}

\revision{%
\section{Proof of Proposition \ref{prop:subspace}}
\label{sec:proof_subspace}

Throughout, $\|\cdot\|$ is the spectral norm.  Write
$\wt{\bs{A}}_i := \wh{\bs{Q}}^{1/2}\bs{A}_i$ and $\bs{E} := \wt{\bs{A}}_i - \bs{\Psi}_i$,
so that $\|\bs{E}\| = e\sigma_1$, where $\sigma_1 \geq \sigma_2 \geq \cdots$ are
the singular values of $\bs{\Psi}_i$, with $\bs{U}_r$ and $\bs{V}_r$ its $r$
principal left and right singular vectors.  Then
$\mu_k = \sigma_k^2/N_{\subsf SRS}$ and
$\delta_r = (\sigma_r^2-\sigma_{r+1}^2)/N_{\subsf SRS}$.  Put
$x := e\mu_1/\delta_r$.  If $x > 1/(2\pi)$ the bound \eqref{eq:sintheta}
exceeds one, so let $x \leq 1/(2\pi) < 1/6$.

Since $\wh{\bs{Q}}_i = \wh{\bs{H}}_i\wh{\bs{H}}_i\herm/N_{\subsf SRS}$ by
\eqref{eq:QHprod}, the matrix $\bs{M}_i$ of \eqref{eq:Mdef} formed from
$\bs{A}_i$ is
\begin{equation} \label{eq:Msub}
    \bs{M}_i \;=\; \wh{\bs{H}}_i\herm\bs{A}_i
    \;=\; \bs{\Psi}_i\herm\wt{\bs{A}}_i
    \;=\; \bs{\Psi}_i\herm\bs{\Psi}_i + \bs{\Psi}_i\herm\bs{E},
\end{equation}
so the $\wbar{N}\times\wbar{N}$ eigenproblem is posed in the whitened
coordinates and no conditioning of $\wh{\bs{Q}}$ enters it.  The implementation
takes $\bs{\Phi}_i$ from the Hermitian part of $\bs{M}_i$, since a finite
\ac{CG} run leaves $\bs{M}_i$ slightly non-Hermitian; that part is
$\bs{\Psi}_i\herm\bs{\Psi}_i + \bs{F}$ with
$\bs{F} := \frac{1}{2}(\bs{\Psi}_i\herm\bs{E} + \bs{E}\herm\bs{\Psi}_i)$ and
$\|\bs{F}\| \leq e\sigma_1^2 = x(\sigma_r^2-\sigma_{r+1}^2)$.  The whitened row
space of the computed projection is $\mc{U} := \mathrm{span}(\wt{\bs{A}}_i\bs{\Phi}_i)$,
and that of \eqref{eq:Giopt} is $\mc{U}^0 := \mathrm{span}(\bs{U}_r)$, the
principal eigenspace of $\bs{\Psi}_i\bs{\Psi}_i\herm = N_{\subsf SRS}\wh{\bs{Q}}_i\wtn$.

\emph{Step 1: the reduced eigenproblem.}\ \ Both $\bs{\Psi}_i\herm\bs{\Psi}_i$ and its perturbation are
Hermitian.  The $r$ largest eigenvalues of the first are at least $\sigma_r^2$
and, by Weyl, the others of the second at most $\sigma_{r+1}^2 + \|\bs{F}\|$, so
the Davis--Kahan $\sin\Theta$ theorem~\cite[Ch.~VII]{bhatia1997matrix} bounds
the largest angle $\theta$ between $\mathrm{span}(\bs{\Phi}_i)$ and
$\mathrm{span}(\bs{V}_r)$:
\begin{equation} \label{eq:dk}
    \sin\theta \;\leq\; \frac{\|\bs{F}\|}{\sigma_r^2-\sigma_{r+1}^2-\|\bs{F}\|}
    \;\leq\; \frac{x}{1-x}.
\end{equation}

\emph{Step 2: the row space.}\ \ With $\bs{P}_\perp := \bs{I}-\bs{U}_r\bs{U}_r\herm$,
$\|\bs{P}_\perp\wt{\bs{A}}_i\bs{\Phi}_i\| \leq \sigma_{r+1}\sin\theta + e\sigma_1$,
and $\wt{\bs{A}}_i\bs{\Phi}_i$ has smallest singular value at least
$\sigma_r\cos\theta - e\sigma_1$.  Since $e\sigma_1 \leq x\sigma_r$ and
$\sigma_{r+1} \leq \sigma_r$,
\begin{equation} \label{eq:dk2}
\begin{aligned}
    \sin\Theta(\mc{U},\mc{U}^0)
    &\leq \frac{\sigma_{r+1}\sin\theta + e\sigma_1}
               {\sigma_r\cos\theta - e\sigma_1}
    \\
    &\leq \frac{x/(1-x) + x}{\sqrt{1-x^2/(1-x)^2}-x}
    \;\leq\; 3x,
\end{aligned}
\end{equation}
the last step for $x < 1/6$.  As $3 < 2\pi$, this is \eqref{eq:sintheta}.
\qed
}

\section{Proof of Proposition \ref{prop:error}}
\label{sec:proof_error}
\revision{%
Throughout, $\|\cdot\|$ is the spectral norm, $\preceq$ the Loewner order, and
the resource element $(n,k)$ is fixed and suppressed.

\emph{Step 1: both projections reduce to a common form.}\ \
Let $\bs{Y}$ denote the approximation matrix.  In both branches it is
Hermitian, as the remainder of the argument requires: in the
polynomial branch $\bs{Y} = \bs{P}^2$ with $\bs{P} = \sum_\ell\beta_\ell
\wh{\bs{Q}}^\ell$ of \eqref{eq:ppoly} a real polynomial in the Hermitian
$\wh{\bs{Q}}$, so $\bs{Y} = \bs{P}\bs{P}\herm \succeq \bs{0}$---a consequence
of that structure, not of squaring \emph{per se}.  In the \ac{CG}
branch, the iterate $\bs{X}^{(t)}$ is not Hermitian in general, and we define the
\ac{CG} approximation used in the analysis, and in the implementation, as the
explicit symmetrization
\begin{equation} \label{eq:Msym}
    \bs{Y} \;:=\; \tfrac{1}{2}\big(\bs{X} + \bs{X}\herm\big),
\end{equation}
which is Hermitian by construction and satisfies
$\|\bs{Y}-\wh{\bs{Q}}^{-1}\| \le \|\bs{X}-\wh{\bs{Q}}^{-1}\|$, since
$\wh{\bs{Q}}^{-1}$ is itself Hermitian and symmetrization is a contraction in
the spectral norm.

With $\bs{Y}$ Hermitian, define
\begin{equation} \label{eq:Sdef}
    \bs{Y}\wtn \;:=\; \bs{Q}^{\subsf H/2}\,\bs{Y}\,\bs{Q}^{1/2},
\end{equation}
a \emph{congruence} of $\bs{Y}$ and hence Hermitian.  It is similar to
$\bs{Q}\bs{Y}$, so $\|\bs{Y}\wtn-\bs{I}\| = \varrho(\bs{Q}\bs{Y}-\bs{I}) \leq
\|\bs{Q}\bs{Y}-\bs{I}\|$ with $\varrho$ the spectral radius, and in the
polynomial branch $\|\bs{Y}\wtn-\bs{I}\| = \|\bs{P}\bs{Q}\bs{P}-\bs{I}\|$
exactly.  In both branches the premise \eqref{eq:cg_premise} therefore gives
\begin{equation} \label{eq:Sbound}
    (1-\epsilon)\bs{I} \;\preceq\; \bs{Y}\wtn \;\preceq\; (1+\epsilon)\bs{I}.
\end{equation}
Because $\bs{Y}\wtn$ is Hermitian, \eqref{eq:Sbound} follows from
$\|\bs{Y}\wtn-\bs{I}\|\le\epsilon$: $\bs{Y}\wtn$ is then unitarily diagonalizable
with real eigenvalues, so that the norm bound confines its spectrum to
$[1-\epsilon,1+\epsilon]$.  The congruence \eqref{eq:Sdef} makes the argument
uniform: under (A2) $\bs{P}^2\bs{Q}$ happens to be
Hermitian, $\bs{P}$ being a polynomial in $\wh{\bs{Q}}=\bs{Q}$, but the
analogous $\bs{Y}\bs{Q}$ of the \ac{CG} branch is not.

Write $\bs{u}_i\wtn := \bs{Q}^{-1/2}\bs{u}_i$ for the whitened direction and
$q_i := \|\bs{u}_i\wtn\|^2 = \bs{u}_i\herm\bs{Q}^{-1}\bs{u}_i$.  By (A1)--(A2)
and Remark~\ref{rem:Gioptfact}, which lets each projection be formed with the
factor $\bs{u}_i$ of $\bs{Q}_i$, the two projections are, up to immaterial scalars
(Lemma~\ref{lem:capupper}),
$\bs{G}_i^0 = \bs{u}_i\herm\bs{Q}^{-1} = \bs{u}_i\wtnh\bs{Q}^{-1/2}$ and
$\bs{G}_i = \bs{u}_i\herm\bs{Y} = \bs{u}_i\wtnh\bs{Y}\wtn\bs{Q}^{-1/2}$: the
ideal projection is the approximate one at $\bs{Y}\wtn=\bs{I}$.

\emph{Step 2: the two \acp{SINR}.}\ \
Put $\xi_1 := \bs{u}_i\wtnh\bs{Y}\wtn\bs{u}_i\wtn$ and
$\xi_2 := \bs{u}_i\wtnh(\bs{Y}\wtn)^2\bs{u}_i\wtn$.  With $\bs{R}_i =
\bs{Q}-\alpha_i\bs{Q}_i$ from \eqref{eq:Qdef}, the projected channel is
$c_i\xi_1$ and the projected interference-plus-noise power
$\xi_2-\alpha_ig_i\xi_1^2$, so \eqref{eq:sinr_proj} gives
\begin{equation} \label{eq:sinr_ideal1}
    \gamma_i = \frac{\alpha_i|c_i|^2}{\xi_2/\xi_1^2-\alpha_ig_i},
    \qquad
    \gamma_i^0 = \frac{\alpha_i|c_i|^2q_i}{1-\alpha_ig_iq_i},
\end{equation}
the second being the first at $\xi_1=\xi_2=q_i$.  Only $|c_i|^2$ varies with
$(n,k)$ and $\Exp|c_i|^2 = g_i$, so
$\Exp(\gamma_i^0) = \alpha_ig_iq_i/(1-\alpha_ig_iq_i)$, where
$\alpha_ig_iq_i<1$ because $\alpha_i\bs{Q}_i \preceq \bs{Q}-\bs{I}$ by
\eqref{eq:Qdef}.

\emph{Step 3: the bound.}\ \
By \eqref{eq:Sbound}, $\xi_1 \geq (1-\epsilon)q_i > 0$ and
$\xi_2 \leq (1+\epsilon)^2q_i$, so $\xi_2/\xi_1^2 \leq \kappa/q_i$ with
$\kappa := (1+\epsilon)^2/(1-\epsilon)^2$.  Since $\gamma_i$ decreases in
$\xi_2/\xi_1^2$,
\begin{equation} \label{eq:kappalam}
    \frac{\gamma_i}{\gamma_i^0} \;\geq\;
    \frac{1-\alpha_ig_iq_i}{\kappa-\alpha_ig_iq_i}
    = \frac{1}{\kappa+(\kappa-1)\Exp(\gamma_i^0)},
\end{equation}
and $(1+\epsilon)^2-(1-\epsilon)^2 = 4\epsilon$ turns the right-hand side into
that of \eqref{eq:sinr_error}.  \qed}

\revision{%
\section{Summary of Notation}
\label{app:notation}

Table~\ref{tab:notation} collects the symbols used throughout the paper.

Four decorations carry meaning.  A \textbf{hat}
denotes a quantity estimated from \ac{SRS} snapshots rather than known
exactly, so $\wh{\bs{Q}}_i$ is the estimate of $\bs{Q}_i$ and $\wh{\bs{H}}_i$
is not an estimate of $\bs{H}_i[n,k]$ at one resource element but the
horizontal concatenation~\eqref{eq:Hhatdef} of the $N_{\subsf SRS}$ sounding
snapshots.  A \textbf{tilde} denotes a quantity after the rank-$r$ projection
$\bs{G}_i$, so $\wt{\bs{H}}_i$ and $\wt{\bs{R}}_i$ are the $r$-dimensional
counterparts of $\bs{H}_i$ and $\bs{R}_i$.  A \textbf{superscript
$\text{\sf w}$} denotes a quantity in the coordinates whitened by
$\bs{Q}^{-1/2}$ ($\wh{\bs{Q}}^{-1/2}$ in Sec.~\ref{sec:subspace}), so
$\bs{Q}_i\wtn = \bs{Q}^{-\subsf H/2}\bs{Q}_i\bs{Q}^{-1/2}$ and
$\bs{G}_i\wtn = \bs{G}_i\bs{Q}^{1/2}$; the whitened coordinates are where
every optimality statement is proved.  A \textbf{parenthesized
superscript} $(t)$ denotes a \ac{CG} iterate at step $t$, so $\bs{X}^{(t)}$ and
the residual and direction blocks belong only to the recursion
\eqref{eq:CG_K}--\eqref{eq:CG_omega}.

Each symbol below has a single meaning; a bold and a non-bold letter,
or one letter with different sub- or superscripts, are different symbols (e.g.,
$\bs{G}_i$ and $G_{\subsf SRS}$).  The one reuse is $k$ as a bound summation index in a
displayed sum over eigen- or Ritz values, where it is local to that sum.

\begin{table}[t]
    \centering
    \caption{\revision{Summary of notation.}}
    \label{tab:notation}
    \scriptsize
    \setlength{\tabcolsep}{4pt}
    \begin{tabular}{c||l}
        \hline
        \textbf{Symbol} & \textbf{Description} \\
        \hline
        \multicolumn{2}{c}{\textit{Dimensions, signals and channels}} \\
        \hline
        $N_{\subsf rx}$ & \acs{BS} receive antennas \\
        $N_{\subsf UE}$ & \acsp{UE} per sector \\
        $N_s$ & Streams per \acs{UE} \\
        $N_{\subsf SRS}$ & \acs{SRS} snapshots per \acs{UE} per $T_{\subsf LT}$ \\
        $\wbar{N}$ & \acs{SRS} columns per \acs{UE}, $N_sN_{\subsf SRS}$ \\
        $r$ & Projection rank \\
        $\bs{y}[n,k]$, $\bs{z}_i[n,k]$ & Received vector; its projection $\bs{G}_i\bs{y}$ \\
        $\bs{H}_i[n,k]$ & Instantaneous channel of \acs{UE} $i$ \\
        $\bs{B}_i$ & A factor of $\bs{Q}_i$ \\
        $\wh{\bs{H}}_i$, $\wh{\bs{H}}_\alpha$ & Concatenated \acs{SRS} estimates of \acs{UE} $i$; all \acsp{UE}, $\alpha$-scaled \eqref{eq:Halpha} \\
        $\wt{\bs{H}}_i[n,k]$ & Projected channel, $\bs{G}_i\bs{H}_i$ \\
        $T_{\subsf LT},T_{\subsf coh},T_{\subsf RE}$ & Long-term, coherence-block and resource-element periods \\
        $\alpha_i$ & Transmit \acs{SNR} of \acs{UE} $i$ per stream \\
        \hline
        \multicolumn{2}{c}{\textit{Covariances, receivers and performance}} \\
        \hline
        $\bs{Q}_i$, $\wh{\bs{Q}}_i$ & Spatial covariance of \acs{UE} $i$; its \acs{SRS} estimate \\
        $\bs{Q}$, $\wh{\bs{Q}}$ & Aggregate $\bs{I}+\sum_j\alpha_j\bs{Q}_j$; its estimate \\
        $\bs{R}_i$, $\wt{\bs{R}}_i$ & Interference-plus-noise covariance; its projection \\
        $\bs{Q}_i\wtn$, $\bs{G}_i\wtn$ & Whitened covariance and projection \\
        $\bs{G}_i$ & Long-term projection of \acs{UE} $i$ \\
        $\bs{F}_i$, $\bs{W}_i$ & Full-dimensional and reduced-domain equalizers \\
        $\bs{M}_i$, $\bs{\Phi}_i$ & Reduced Gram matrix; its principal eigenvectors \\
        $\bs{A}_i$, $\bs{\Psi}_i$ & \acs{CG} solution block; whitened \acs{SRS} block \\
        $\bs{X}$, $\bs{Y}$ & \acs{CG} approximation of $\wh{\bs{Q}}^{-1}$; its symmetrization \\
        $\bs{P}(\bs{\beta})$, $\bs{\beta}$ & Polynomial approximant of $\wh{\bs{Q}}^{-1/2}$; its coefficients \\
        $C_i$, $\wbar{C}_i$ & Ergodic capacity of the projected channel; Jensen bound \\
        $\gamma_i$, $\Lambda_i$ & Post-projection \acs{SINR}; signal-to-interference-ratio matrix \\
        $\mu_k,\delta_r$; $e,\epsilon,\Theta$ & Ritz values of $\bs{Q}_i\wtn$ and gap; solve, operator and angle error \\
        \hline
    \end{tabular}
\end{table}
}

\bibliographystyle{refs/IEEEtran}
\bibliography{refs/IEEEfull,refs/bibliography}{}

@article{kang2024cellular,
  title={{Cellular wireless networks in the upper mid-band}},
  author={Kang, Seongjoon and Mezzavilla, Marco and Rangan, Sundeep and Madanayake, Arjuna and Venkatakrishnan, Satheesh Bojja and Hellbourg, Gr{\'e}gory and Ghosh, Monisha and Rahmani, Hamed and Dhananjay, Aditya},
  journal={IEEE Open Journal of the Communications Society},
  volume={5},
  pages={2058--2075},
  year={2024},
  publisher={IEEE}
}

@article{jeon2019decentralized,
  title={{Decentralized equalization with feedforward architectures for massive MU-MIMO}},
  author={Jeon, Charles and Li, Kaipeng and Cavallaro, Joseph R and Studer, Christoph},
  journal={IEEE Transactions on Signal Processing},
  volume={67},
  number={17},
  pages={4418--4432},
  year={2019},
  publisher={IEEE}
}

@book{heath2018foundations,
  title={{Foundations of MIMO communication}},
  author={Heath Jr, Robert W and Lozano, Angel},
  year={2018},
  publisher={Cambridge University Press}
}

@inproceedings{yin2015vlsi,
  title={{VLSI design of large-scale soft-output MIMO detection using conjugate gradients}},
  author={Yin, Bei and Wu, Michael and Cavallaro, Joseph R and Studer, Christoph},
  booktitle={2015 IEEE International Symposium on Circuits and Systems (ISCAS)},
  pages={1498--1501},
  year={2015},
  organization={IEEE}
}

@inproceedings{lozano2007long,
  title={Long-term transmit beamforming for wireless multicasting},
  author={Lozano, Angel},
  booktitle={2007 IEEE International Conference on Acoustics, Speech and Signal Processing-ICASSP'07},
  volume={3},
  pages={III--417},
  year={2007},
  organization={IEEE}
}

@article{yu2020long,
  title={{Long-term channel statistic estimation for highly-mobile hybrid mmWave multi-user MIMO systems}},
  author={Yu, Jiadong and Liu, Xiaolan and Qi, Haoran and Gao, Yue},
  journal={IEEE Transactions on Vehicular Technology},
  volume={69},
  number={12},
  pages={14277--14289},
  year={2020},
  publisher={IEEE}
}

@inproceedings{dai2021scalable,
  title={A scalable generator for massive MIMO baseband processing systems with beamspace channel estimation},
  author={Dai, Yue and Liew, Harrison and Rasekh, Maryam Eslami and Mirfarshbafan, Seyed Hadi and Gallyas-Sanhueza, Alexandra and Dunn, James and Madhow, Upamanyu and Studer, Christoph and Nikoli{\'c}, Borivoje},
  booktitle={2021 IEEE Workshop on Signal Processing Systems (SiPS)},
  pages={182--187},
  year={2021},
  organization={IEEE}
}

@book{marzetta2016fundamentals,
  title={Fundamentals of massive MIMO},
  author={Marzetta, Thomas L and Larsson, Erik G and Yang, Hong and Ngo, Hien Quoc},
  year={2016},
  publisher={Cambridge University Press}
}

@article{lin2019hybrid,
  title={Hybrid beamforming for millimeter wave systems using the MMSE criterion},
  author={Lin, Tian and Cong, Jiaqi and Zhu, Yu and Zhang, Jun and Letaief, Khaled Ben},
  journal={IEEE Transactions on Communications},
  volume={67},
  number={5},
  pages={3693--3708},
  year={2019},
  publisher={IEEE}
}

@article{visotsky2002space,
  title={Space-time transmit precoding with imperfect feedback},
  author={Visotsky, Eugene and Madhow, Upamanyu},
  journal={IEEE transactions on Information Theory},
  volume={47},
  number={6},
  pages={2632--2639},
  year={2002},
  publisher={IEEE}
}

@article{jafar2004transmitter,
  title={Transmitter optimization and optimality of beamforming for multiple antenna systems},
  author={Jafar, Syed Ali and Goldsmith, Andrea},
  journal={IEEE Transactions on Wireless Communications},
  volume={3},
  number={4},
  pages={1165--1175},
  year={2004},
  publisher={IEEE}
}

@article{mirfarshbafan2020beamspace,
  title={Beamspace channel estimation for massive MIMO mmWave systems: Algorithm and VLSI design},
  author={Mirfarshbafan, Seyed Hadi and Gallyas-Sanhueza, Alexandra and Ghods, Ramina and Studer, Christoph},
  journal={IEEE Transactions on Circuits and Systems I: Regular Papers},
  volume={67},
  number={12},
  pages={5482--5495},
  year={2020},
  publisher={IEEE}
}

@inproceedings{sayeed2013beamspace,
  title={Beamspace MIMO for high-dimensional multiuser communication at millimeter-wave frequencies},
  author={Sayeed, Akbar and Brady, John},
  booktitle={2013 IEEE global communications conference (GLOBECOM)},
  pages={3679--3684},
  year={2013},
  organization={IEEE}
}

@inproceedings{wu2013approximate,
  title={Approximate matrix inversion for high-throughput data detection in the large-scale MIMO uplink},
  author={Wu, Michael and Yin, Bei and Vosoughi, Aida and Studer, Christoph and Cavallaro, Joseph R and Dick, Chris},
  booktitle={2013 IEEE international symposium on circuits and systems (ISCAS)},
  pages={2155--2158},
  year={2013},
  organization={IEEE}
}

@article{kammoun2014linear,
  title={Linear precoding based on polynomial expansion: Large-scale multi-cell MIMO systems},
  author={Kammoun, Abla and M{\"u}ller, Axel and Bj{\"o}rnson, Emil and Debbah, M{\'e}rouane},
  journal={IEEE Journal of Selected Topics in Signal Processing},
  volume={8},
  number={5},
  pages={861--875},
  year={2014},
  publisher={IEEE}
}

@book{powell1981approximation,
  title={Approximation theory and methods},
  author={Powell, Michael James David},
  year={1981},
  publisher={Cambridge university press}
}

@book{horn2012matrix,
  title={Matrix analysis},
  author={Horn, Roger A and Johnson, Charles R},
  year={2012},
  publisher={Cambridge university press}
}

@techreport{nokia2025massiveMIMO,
  title        = {{Extreme Massive MIMO for Macro Cell Capacity Boost in 5G-Advanced and 6G}},
  author       = {Harri Holma, Harish Viswanathan and Preben Mogensen},
  year         = {2025},
  institution  = {Nokia},
  type         = {White Paper},
  url          = {https://www.nokia.com/asset/210786/}
}

@techreport{nokia2025coverage7to15GHz,
  title        = {Coverage Evaluation of 7–15 GHz Bands from Existing Sites},
  author       = {{Nokia}},
  year         = {2025},
  institution  = {Nokia},
  type         = {White Paper},
  url          = {https://www.nokia.com/asset/213702/},
  note         = {Accessed October 30, 2025}
}

@article{jia2025joint,
  title={{Joint Detection, Channel Estimation and Interference Nulling for Terrestrial-Satellite Downlink Co-Existence in the Upper Mid-Band}},
  author={Jia, Shizhen and Ying, Mingjun and Mezzavilla, Marco and Calin, Doru and Rappaport, Theodore S and Rangan, Sundeep},
  journal={arXiv preprint arXiv:2510.08824},
  year={2025}
}

@article{marzetta2002capacity,
  title={{Capacity of a mobile multiple-antenna communication link in Rayleigh flat fading}},
  author={Marzetta, Thomas L. and Hochwald, Bertrand M.},
  journal={IEEE transactions on Information Theory},
  volume={45},
  number={1},
  pages={139--157},
  year={2002},
  publisher={IEEE}
}

@article{rasteh2025spatial,
  title={A Spatial Array for Spectrally Agile Wireless Processing},
  author={Rasteh, Ali and Hennessee, Andrew and Shivhare, Ishaan and Garg, Siddharth and Rangan, Sundeep and Reagen, Brandon},
  journal={arXiv preprint arXiv:2512.04182},
  year={2025}
}

@inproceedings{mezzavilla2025prototyping,
  title={{6G} Prototyping in the Upper Mid-Band (7--24 {GHz})},
  author={Mezzavilla, Marco and Rasteh, Ali and Zappe, Michael and Zappe, Elijah and Dhananjay, Aditya and Rangan, Sundeep},
  booktitle={2025 IEEE Wireless Communications and Networking Conference (WCNC)},
  pages={1--3},
  year={2025},
  doi={10.1109/WCNC61545.2025.10978617}
}

@inproceedings{rasteh2024nearfield,
  title={Near-Field Measurement System for the Upper Mid-Band},
  author={Rasteh, Ali and Hari, Raghavendra Palayam and Guo, Hao and Mezzavilla, Marco and Rangan, Sundeep},
  booktitle={2024 58th Asilomar Conference on Signals, Systems, and Computers},
  pages={1684--1689},
  year={2024},
  doi={10.1109/IEEECONF60004.2024.10942861}
}

@inproceedings{tian2024flud,
  title={FLUD: A Scalable and Configurable Systolic Array Design for LU Decomposition on FPGAs},
  author={Tian, Xingyu and Yang, Geng and Fang, Zhenman},
  booktitle={2024 International Conference on Field Programmable Technology (ICFPT)},
  pages={01--09},
  year={2024},
  organization={IEEE}
}

@article{he2020model,
  title={Model-driven deep learning for massive multiuser MIMO constant envelope precoding},
  author={He, Yunfeng and He, Hengtao and Wen, Chao-Kai and Jin, Shi},
  journal={IEEE Wireless Communications Letters},
  volume={9},
  number={11},
  pages={1835--1839},
  year={2020},
  publisher={IEEE}
}

@article{hu2020iterative,
  title={Iterative algorithm induced deep-unfolding neural networks: Precoding design for multiuser MIMO systems},
  author={Hu, Qiyu and Cai, Yunlong and Shi, Qingjiang and Xu, Kaidi and Yu, Guanding and Ding, Zhi},
  journal={IEEE Transactions on Wireless Communications},
  volume={20},
  number={2},
  pages={1394--1410},
  year={2020},
  publisher={IEEE}
}

@inproceedings{hoydis2023sionna,
  title={Sionna RT: Differentiable ray tracing for radio propagation modeling},
  author={Hoydis, Jakob and A{\"\i}t Aoudia, Fay{\c{c}}al and Cammerer, Sebastian and Nimier-David, Merlin and Binder, Nikolaus and Marcus, Guillermo and Keller, Alexander},
  booktitle={2023 IEEE Globecom Workshops (GC Wkshps)},
  pages={317--321},
  year={2023},
  organization={IEEE}
}

@article{o1980block,
  title={The block conjugate gradient algorithm and related methods},
  author={O'Leary, Dianne P},
  journal={Linear algebra and its applications},
  volume={29},
  pages={293--322},
  year={1980},
  publisher={Elsevier}
}

@book{nocedal2006numerical,
  title={Numerical optimization},
  author={Nocedal, Jorge and Wright, Stephen J},
  year={2006},
  publisher={Springer}
}

@book{bhatia1997matrix,
  title={Matrix Analysis},
  author={Bhatia, Rajendra},
  series={Graduate Texts in Mathematics},
  volume={169},
  year={1997},
  publisher={Springer-Verlag, New York}
}

@article{lu2024nearfield,
  title={A tutorial on near-field {XL-MIMO} communications toward {6G}},
  author={Lu, Haiquan and Zeng, Yong and You, Changsheng and Han, Yu and Zhang, Jiayi and Wang, Zhe and Dong, Zhenjun and Jin, Shi and Wang, Cheng-Xiang and Jiang, Tao and You, Xiaohu and Zhang, Rui},
  journal={IEEE Communications Surveys \& Tutorials},
  volume={26},
  number={4},
  pages={2213--2257},
  year={2024},
  publisher={IEEE}
}

@article{wang2024xlmimo,
  title={Extremely large-scale {MIMO}: Fundamentals, challenges, solutions, and future directions},
  author={Wang, Zhe and Zhang, Jiayi and Du, Hongyang and Sha, Wei E. I. and Ai, Bo and Niyato, Dusit and Debbah, M{\'e}rouane},
  journal={IEEE Wireless Communications},
  volume={31},
  number={3},
  pages={117--124},
  year={2024},
  publisher={IEEE}
}

@article{carvalho2020nonstationarities,
  title={Non-stationarities in extra-large-scale massive {MIMO}},
  author={de Carvalho, Elisabeth and Ali, Anum and Amiri, Abolfazl and Angjelichinoski, Marko and Heath, Robert W.},
  journal={IEEE Wireless Communications},
  volume={27},
  number={4},
  pages={74--80},
  year={2020},
  publisher={IEEE}
}

@inproceedings{amiri2018extremely,
  title={Extremely large aperture massive {MIMO}: Low complexity receiver architectures},
  author={Amiri, Abolfazl and Angjelichinoski, Marko and de Carvalho, Elisabeth and Heath, Robert W.},
  booktitle={IEEE Globecom Workshops (GC Wkshps)},
  pages={1--6},
  year={2018},
  organization={IEEE}
}

@article{cui2022channel,
  title={Channel estimation for extremely large-scale {MIMO}: Far-field or near-field?},
  author={Cui, Mingyao and Dai, Linglong},
  journal={IEEE Transactions on Communications},
  volume={70},
  number={4},
  pages={2663--2677},
  year={2022},
  publisher={IEEE}
}

@article{hu2018beyond,
  title={Beyond massive {MIMO}: The potential of data transmission with large intelligent surfaces},
  author={Hu, Sha and Rusek, Fredrik and Edfors, Ove},
  journal={IEEE Transactions on Signal Processing},
  volume={66},
  number={10},
  pages={2746--2758},
  year={2018},
  publisher={IEEE}
}

@article{bjornson2020power,
  title={Power scaling laws and near-field behaviors of massive {MIMO} and intelligent reflecting surfaces},
  author={Bj{\"o}rnson, Emil and Sanguinetti, Luca},
  journal={IEEE Open Journal of the Communications Society},
  volume={1},
  pages={1306--1324},
  year={2020},
  publisher={IEEE}
}

@article{akyildiz2016realizing,
  title={Realizing ultra-massive {MIMO} $(1024\times1024)$ communication in the $(0.06$--$10)$ terahertz band},
  author={Akyildiz, Ian F. and Jornet, Josep Miquel},
  journal={Nano Communication Networks},
  volume={8},
  pages={46--54},
  year={2016},
  publisher={Elsevier}
}

@article{faisal2020ultramassive,
  title={Ultramassive {MIMO} systems at terahertz bands: Prospects and challenges},
  author={Faisal, Alice and Sarieddeen, Hadi and Dahrouj, Hayssam and Al-Naffouri, Tareq Y. and Alouini, Mohamed-Slim},
  journal={IEEE Vehicular Technology Magazine},
  volume={15},
  number={4},
  pages={33--42},
  year={2020},
  publisher={IEEE}
}

@article{sarieddeen2021overview,
  title={An overview of signal processing techniques for terahertz communications},
  author={Sarieddeen, Hadi and Alouini, Mohamed-Slim and Al-Naffouri, Tareq Y.},
  journal={Proceedings of the IEEE},
  volume={109},
  number={10},
  pages={1628--1665},
  year={2021},
  publisher={IEEE}
}

@article{bjornson2025gigantic,
  title={Enabling {6G} performance in the upper mid-band by transitioning from massive to gigantic {MIMO}},
  author={Bj{\"o}rnson, Emil and Kara, Ferdi and Kolomvakis, Nikolaos and Kosasih, Alva and Ramezani, Parisa and Salman, Murat Babek},
  journal={IEEE Open Journal of the Communications Society},
  volume={6},
  pages={5450--5463},
  year={2025},
  publisher={IEEE}
}

@article{bjornson2019massive,
  title={Massive {MIMO} is a reality --- What is next? Five promising research directions for antenna arrays},
  author={Bj{\"o}rnson, Emil and Sanguinetti, Luca and Wymeersch, Henk and Hoydis, Jakob and Marzetta, Thomas L.},
  journal={Digital Signal Processing},
  volume={94},
  pages={3--20},
  year={2019},
  publisher={Elsevier}
}

@article{adhikary2013joint,
  title={Joint spatial division and multiplexing---The large-scale array regime},
  author={Adhikary, Ansuman and Nam, Junyoung and Ahn, Jae-Young and Caire, Giuseppe},
  journal={IEEE Transactions on Information Theory},
  volume={59},
  number={10},
  pages={6441--6463},
  year={2013},
  publisher={IEEE}
}

@article{nam2014joint,
  title={Joint spatial division and multiplexing: Opportunistic beamforming, user grouping and simplified downlink scheduling},
  author={Nam, Junyoung and Adhikary, Ansuman and Ahn, Jae-Young and Caire, Giuseppe},
  journal={IEEE Journal of Selected Topics in Signal Processing},
  volume={8},
  number={5},
  pages={876--890},
  year={2014},
  publisher={IEEE}
}

@techreport{3gpp38211,
  title={{NR}; Physical channels and modulation},
  author={{3GPP}},
  institution={3rd Generation Partnership Project},
  type={Technical Specification},
  number={TS 38.211},
  note={V18.6.0, Release 18},
  year={2025}
}

@article{kiani2026interference,
  title={Interference Suppression for Massive {MU-MIMO} Long-Term Beamforming with Matrix Inversion Approximation},
  author={Kiani, Amirreza and Rasteh, Ali and Mezzavilla, Marco and Rangan, Sundeep},
  journal={arXiv preprint arXiv:2604.06355},
  year={2026}
}

@article{kiani2026lowrank,
  title={Low-rank Preconditioning in Beamspace Domain for Massive {MU-MIMO} Long-Term Beamforming},
  author={Kiani, Amirreza and Rasteh, Ali and Mezzavilla, Marco and Rangan, Sundeep},
  journal={arXiv preprint arXiv:2605.02748},
  year={2026}
}


\end{document}